\documentclass[10pt]{article}
\usepackage{ametsoc2col}
\usepackage{overpic}
\usepackage{amssymb}
\usepackage{amsmath}
\usepackage{subeqnarray}
\numberwithin{equation}{section}
\usepackage{amsfonts}
\usepackage{amsxtra}
\usepackage{enumerate}

\newcommand{\ttrun}{t_{\rm run}}
\newcommand{\percent}{\%}
\newcommand{\bu}{\mathbf{u}}
\newcommand{\Umax}{U_{\rm max}}
\newcommand{\rr}{r}
\newcommand{\psideep}{\bar{\psi}_{\rm deep}}
\newcommand{\psideeptilde}{\tilde\psi_{\rm deep}}
  \newcommand{\uudeep}{\bar{u}_{\rm deep}}
\newcommand{\uu}{u}
\newcommand{\vv}{v}
\newcommand{\qq}{q}
\newcommand{\forcing}{F}
\newcommand{\dissipation}{D}
\newcommand{\forcingbar}{\bar{\forcing}}
\newcommand{\acforcing}{{A}}
\newcommand{\acforcingmax}{{A_{\rm  max}}}
\newcommand{\zetabar}{\bar{\zeta}}

\newcommand{\bx}{\mathbf{x}}
\newcommand{\bxc}{\mathbf{x}_{\rm c}}
\newcommand{\xc}{x_{\rm c}}
\newcommand{\yc}{y_{\rm c}}
\newcommand{\ttmax}{t_{\rm max}}
\newcommand{\Deltax}{\Delta x}
\newcommand{\Deltat}{\Delta t}
\newcommand{\ramp}{\rho}
\newcommand{\psilim}{\psi_{\rm{lim}}}
\newcommand{\s}{s}
\newcommand{\cc}{c}

\newcommand{\qqbar}{\bar{\qq}}
\newcommand{\qqmax}{\qq_{\rm max}}
\newcommand{\qqstarmax}{\qq^*_{\rm max}}
\newcommand{\qqpeak}{\qq_{\rm pk}}
\newcommand{\qqpeakc}{\qq_{\rm pkC}}
\newcommand{\qqpeaka}{\qq_{\rm pkA}}
\newcommand{\qqpatch}{\qq_{\rm patch}}
\newcommand{\qqstarpatch}{\qq^*_{\rm patch}}
\newcommand{\kkmin}{k_{\rm min}}
\newcommand{\ttrel}{t_{\rm rel}}

\newcommand{\uubar}{\bar{\uu}}
\newcommand{\vvbar}{\bar{\vv}}
\newcommand{\psibar}{\bar{\psi}}
\newcommand{\fractionalbias}{b}
\newcommand{\fractionalbiasmax}{b_{\rm max}}
\newcommand{\oneminusbias}{B}

\newcommand{\oneandahalflayer}{\mbox{$1\Antisliver\half$-layer}}
\newcommand{\qg}{quasi-geo\-stro\-phic}
\newcommand{\qf}{qua\-si\-fric\-tion}
\newcommand{\Qf}{Qua\-si\-fric\-tion}
\newcommand{\wl} {wea\-ther layer}
\newcommand{\whl}{wea\-ther-layer}

\newcommand{\andrefs}{{\& ref\-er\-enc\-es there\-in}}
\newcommand{\LD}{L_{\rm D}}
\newcommand{\kd}{k_{\rm D}}

\newcommand{\sliver}{{\hskip0.6pt}}

\newcommand{\Sliver}{{\hskip1.2pt}}
\newcommand{\antisliver}{{\hskip-0.6pt}}

\newcommand{\Antisliver}{{\hskip-1.2pt}}

\newcommand{\definedas}{\,:=\,}
\newcommand{\sfrac}[2]{{\textstyle\frac{#1}{#2}}}

\newcommand{\half}   {\sfrac{1}{2}}

\newcommand{\degree}{{$^\circ$}}
\newcommand{\yr}{{\sliver yr}}
\newcommand{\km}{{\Sliver km}}
\newcommand{\m}{{\Sliver m}}
\newcommand{\ms}{{\Sliver m\Sliver s$^{-1}$}}
\newcommand{\mss}{{\Sliver m\Sliver s$^{-2}$}}
\newcommand{\mms}{{\Sliver m$^2$s$^{-1}$}}
\newcommand{\mmss}{{\Sliver m$^2$s$^{-2}$}}
\newcommand{\smone}{{\Sliver s$^{-1}$}}
\newcommand{\smmone}{{\Sliver s$^{-1}$m$^{-1}$}}

\newcommand{\wkg}{{\Sliver W\sliver kg$^{-1}$}}
\newcommand{\jkg}{{\Sliver J\sliver kg$^{-1}$}}

\newcommand{\myabstract}{
A longstanding mystery about Jupiter has been the straightness
and steadiness of its weather-layer jets, quite unlike
terrestrial strong jets with their characteristic unsteadiness
and long-wavelength meandering.  The problem is addressed in
two steps.  The first is to take seriously the classic
Dowling-Ingersoll (DI) \oneandahalflayer\ scenario and its
supporting observational evidence, pointing toward
deep, massive, zonally-symmetric zonal jets in
the underlying dry-convective layer.  The second
is to improve the realism of the model stochastic forcing
used to represent the effects of Jupiter's moist convection
as far as possible within the \oneandahalflayer\ dynamics.  The
real moist convection should be strongest in the belts where
the interface to the deep flow is highest and coldest, and
should generate cyclones as well as anticyclones with the
anticyclones systematically stronger.  The new model forcing
reflects these insights.  Also, it acts \qf ally on
large scales to produce statistically steady turbulent
weather-layer regimes without any need for explicit
large-scale dissipation, and with
\whl\
jets that are approximately straight
thanks to
the influence of the deep jets,
allowing
shear stability despite
nonmonotonic potential-vorticity gradients
when the Rossby deformation lengthscale is not too large.
Moderately strong forcing produces
chaotic vortex dynamics and realistic belt-zone contrasts in
the model's convective activity, through an eddy-induced
sharpening and strengthening of the \whl\ jets relative to the
deep jets, tilting the interface between them.
Weak forcing, for which the
only jet-sharpening mechanism is the passive (Kelvin) shearing
of vortices
(as in the ``CE2'' or ``SSST'' theories),
produces unrealistic belt-zone contrasts.
}

\begin{document}
\title{\textbf{\large{Jupiter's unearthly jets:
    a new turbulent
    model exhibiting\\
    statistical steadiness 
    without large-scale dissipation}}}
    \author{\textsc{Stephen I. Thomson$^*$,
Michael E. McIntyre}\\
\textit{\footnotesize{Department of Applied Mathematics and 
Theoretical Physics, University of Cambridge, CB3 0WA, UK}}\\
\textit{\footnotesize{$^*$Present affiliation:
Dept.\ of Mathematics,
University of Exeter, North Park Road, Exeter EX4 4QF, UK}}\\
{\footnotesize{emps.exeter.ac.uk/mathematics/staff/sit204/,
stephen.i.thomson@gmail.com; \
www.damtp.cam.ac.uk/user/mem2/,
mem2@cam.ac.uk}}\\
\footnotesize{Submitted to \textit{J. Atmos.\ Sci.} 9 Dec 2014,
revised 15 Jul \& 23   
Sept 2015}
  \vspace {-0.4cm}
}
\ifthenelse{\boolean{dc}}
{
\twocolumn[
\begin{@twocolumnfalse}
\amstitle

\begin{center}
\begin{minipage}{13.0cm}
\begin{abstract}
  \myabstract
  \newline
  \begin{center}
    \rule{38mm}{0.2mm}
  \end{center}
\end{abstract}
\end{minipage}
\end{center}
\end{@twocolumnfalse}
]
}
{         
\amstitle
\begin{abstract}
\myabstract
\end{abstract}
\newpage
}

\section{Introduction}
\label{sec:intro}

The
aim of this work is to find the simplest
stochastically forced model of Jupiter's visible \wl\ that
reproduces the straightness and steadiness of the observed
prograde jets,
and the belt--zone contrasts in small-scale convective activity,
under a forcing regime that is
arguably closer to the real planet's than either
(a) the forcing used
in orthodox beta-turbulence models, or
(b) the purely anticyclonic forcing used
in the recent work of \citet{Li2006} and \citet{Showman2007}.
The new forcing is discussed below, after sketching the
scientific background.

The \wl\ is the cloudy
moist-convective
layer overlying a much deeper, hotter dry-convective layer.
Such vertical structure, though not directly observed,
is to be expected from
the need to carry a substantial heat flux from below
and from the basic thermodynamics and estimated
chemical composition of Jupiter's atmosphere
\citep[e.g.,][\andrefs]{Sugiyama2006}.

Even at high latitudes, the observed
\whl\ jets are ``straight'' in the sense that
apart from small-scale disturbances they
closely follow parallels of latitude,
especially the prograde jets,
as
clearly seen in
the well-known animated polar view from Cassini
images.\footnote{A
movie is available from
www.atm.damtp.cam.ac.uk/people/  \break
/mem/nasa-cassini-index.html, by kind courtesy of
NASA and the Cassini mission
\citep[][movie S1]{Porco2003}.
  }
This extreme straightness
contrasts with the
meandering behavior
found in
many single-layer model studies
including the work of
\citet{Li2006} and \citet{Showman2007} and, for instance,
\citet{Scott2007}.
Jupiter's jets are also remarkably close to being steady, as
evidenced by the almost identical zonal-mean zonal wind profiles
seen in 1979 and 2000, from cloud tracking in the Voyager 1 and
Cassini images \citep{Limaye1986,Porco2003}.

Because the \wl\ appears turbulent
and has no solid lower boundary, we
confine attention to models that can reach
statistically steady states
without the large-scale friction used in the
beta-turbulence models.
We also avoid the use of
large-scale Newtonian cooling as an eddy-damping
mechanism.
Real radiative heat transfer is not only
far more complicated, but also
dependent on unknown details of the
cloud structure within the
\wl\
and near the interface with the dry-convective layer.

The extreme
straightness and steadiness
of Jupiter's prograde jets make them
strikingly dissimilar to the strong jets
of the Earth's atmosphere and oceans,
with their
conspicuous, large-amplitude
long wave meandering.
By strong jets we mean
the atmosphere's tropopause and polar-night jets,
and the strongest ocean currents
such as the Gulf Stream, the Kuroshio and the Agulhas.
The cores of these
terrestrial strong jets are marked by concentrated
isentropic gradients of Rossby--Ertel
potential vorticity
or gradients of ocean surface temperature,
inversion of which implies
sharp velocity profiles
having width scales of the order of
an appropriate Rossby deformation
lengthscale
$\LD$.
Such meandering strong jets
are quite different from the
straighter but very weak topography-constrained
``ghost'' or ``latent'' jets
in the Pacific ocean, visible only
after much time-averaging
\citep[][\andrefs]{Maximenko2005}.

Jupiter's jets
are hardly weak.  On the contrary,
at least some of them are
strong enough to look shear-unstable, by some criteria, with
nonmonotonic potential-vorticity gradients
\citep[e.g.,][]{Dowling1989,Read2006,Marcus2011a}.
Here we argue that
their straightness and steadiness
may come from a different,
strictly extraterrestrial
combination of circumstances, which
calls for significant modeling innovations.

We propose a new idealized model whose
two most crucial aspects are as follows.  The first is the
stochastic forcing
of turbulence by thunderstorms
and other small-scale moist-convective elements
injected into the \wl\
from the underlying dry-convective layer.\footnote{We
deliberately exclude
other excitation mechanisms.  In particular, we exclude
terrestrial-type baroclinic instabilities.
These are arguably weak or absent because of the
absence
of a solid lower surface at the base of the \wl,
and because of the weak
pole-to-equator temperature gradient --- a weakness
expected, in turn, from the well known
``convective thermostat'' argument
\citep{Ingersoll1978}.
  }
We assume that moist convection generates small cyclones and
anticyclones,
with a bias toward stronger
anticylones.
Such a ``potential-vorticity bias'' or ``PV bias''
recognizes that heat as
well as mass is injected.  This contrasts with the mass-only,
anticyclones-only scenarios of \citet{Li2006} and
\citet{Showman2007} on the one hand, and with the
perfectly unbiased small-scale forcing
used in
beta-turbulence models
on the other.  PV bias will
enable us to dispense with the large-scale friction
required for statistical steadiness
in beta-turbulence models.

The second aspect is the presence of zonally-symmetric
deep zonal jets in the underlying dry-convective layer.
They will prove crucial to our model's behavior.
Here we follow the pioneering work of
\citet[][hereafter DI]{Dowling1989},
who produced cloud-wind evidence pointing to
two remarkable and surprising conclusions.  The first is
that the large-scale vortex dynamics,
in latitudes around 15\degree--35\degree\ at least,
is approximately the same as
the dynamics of a potential-vorticity-conserving
\oneandahalflayer\ model, with the upper layer
representing the entire \wl.
DI's second conclusion is that the cloud-wind data
can be fitted into this picture only if the underlying
dry-convective layer is in large-scale relative motion.
The simplest possibility allowing a good fit is that the
relative motion consists of deep zonally-symmetric zonal
jets.  Those deep jets must have substantial velocities,
comparable in
order of magnitude to jet velocities at cloud-top levels.
To our knowledge, no subsequent cloud-wind study has
overturned this second conclusion.
So we use a \oneandahalflayer\ model with deep jets.  We treat
the deep jets as prescribed and steady,
consistent
with the far greater
depth and mass of the dry-convective layer.

The relevance of \oneandahalflayer\ dynamics has
recently gained support from a different direction.
\citet[]
[]{Sugiyama2014}
present results
from a two-dimen\-sion\-al cloud resolving mo\-del
that includes the condensation
and precipitation of water and other minor species.
A model \wl\ emerges whose stable stratification
is sharply concentrated near the interface
with the dry-convective layer,
as a consequence of water-cloud behavior.
This result suggests that
the real \wl\ could be surprisingly close to the
\oneandahalflayer\ idealization with its perfectly sharp
interface.  Further such work is needed,
if only because
the real cloud-scale moist convection must
be three-dimensional, as suggested by the morphology both of
Jupiter's folded filamentary regions and of terrestrial
supercell or ``tornado alley''
thunderstorms, with their intertwined patterns of updrafts,
downdrafts, and precipitation.

DI's evidence for deep jets
remains important today
because, as yet, there are no other observational
constraints on the existence or nonexistence of the deep jets,
outside the equatorial region.
No such constraints are expected until, hopefully,
gravitational data come in from the Juno mission in 2016.
Numerical studies of the dry-convective layer
cannot address the question
because they
need to make speculative assumptions
about conditions at depth,
including
the effective bottom boundary conditions felt by
Taylor-Proudman-constrained
deep jets at latitudes within the
associated tangent cylinder.
Here there is great uncertainty.
There is a range of possible conditions
whose extremes are
a slippery radiative layer well above the
metallic-hydrogen transition,
making deep jets easy to generate, and
a no-slip magnetohydrodynamic transition layer that
inhibits them
\citep[e.g.,][\andrefs]{Guillot2005a,Jones2009,Liu2013,
Gastine2014}.
Jupiter's prograde equatorial jet system
needs separate consideration,
being almost certainly
outside any relevant tangent cylinder or cylinders.

Zonal symmetry or straightness
is plausible
for any deep, dry-convective jets that may exist,
in virtue of the scale separation between the jets themselves
and the relatively tiny, Coriolis-constrained convective
elements that excite them, as seen in the numerical studies
\citep[e.g.,][\andrefs]{Jones2009,Gastine2014}.

Our own relatively modest aim, then, is to see whether,
on the basis of the DI scenario with prescribed
deep, straight jets, and nonmonotonic upper PV gradients,
an idealized \oneandahalflayer\ model
with PV-biased forcing
can produce not only
statistical steadiness
in the absence of large scale dis\-si\-pa\-tion
but also realistic,
quasi-zonal
large scale \whl\
structures, with moist-convective forcing strongest
in the cyclon\-ically-sheared ``belts'' and
weakest in the anti\-cy\-clon\-ically-sheared ``zones''.
The folded filamentary regions and lightning
observed on the real planet,
assumed to be symptomatic of moist convection,
are
concentrated in the belts \citep[e.g.,][]{Porco2003}.
In addition we aim
to test the effectiveness, within the
idealized model, of the
beta-drift-mediated migration of
small anticyclones from belts into zones, following a
suggestion by \citet{Ingersoll2000} that such migration
might be significant.

\begin{figure} [t]
\includegraphics[width=21.3pc,angle=0]
{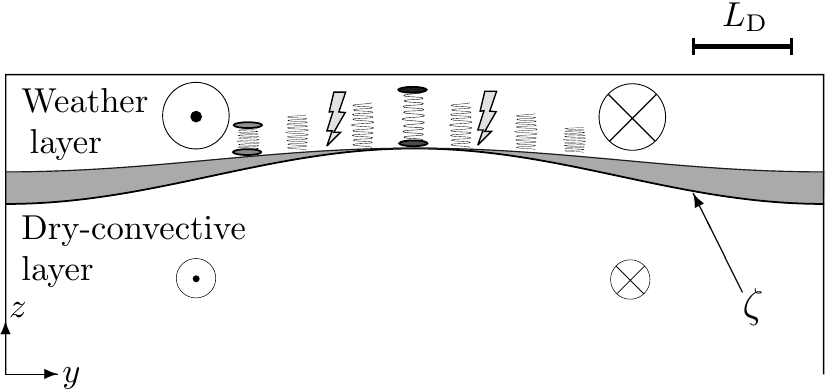}
\caption{\footnotesize
Schematic of the model setup and motivation;
see text.  The bar at top right indicates
an $\LD$ value well below the threshold for
marginal shear instability
with nonmonotonic upper PV gradients.
The notional
cumulonimbus clouds,
concentrated in the model belt,
can be thought of
as tending to generate
vortex pairs with cyclones below and anticyclones above.
Such vortex pairs, called
``hetons'' or ``heatons''
in the oceanographic literature, can tilt and then propagate like
ordinary two-dimensional vortex pairs. 
Ordinary vortex pairs
are all that can be accommodated in
a \oneandahalflayer\ model.
\citet{Ingersoll2000} remind us that ``both cyclonic and
anticyclonic structures exist within the belts'' of the
real planet, and succinctly
summarize the case for their being generated by moist convection.
  }
\label{fig:cartoon}
\end{figure}

We also
look at the issue of
shear instability
and the relevance of
Arnol'd's second stability criterion for
nonmonotonic PV gradients
(``A2 stability''), following
suggestions by \citet{Dowling1993} and \citet{Stamp1993}.
Jet straightness implies that the \wl\ is either marginal or
submarginal to shear instability; but the model results
will force us to a stronger, and we believe
novel, conclusion,
differing from Dowling's original suggestion, namely
that the \wl\ must be well
\emph{below} the threshold for
marginal instability.
In the model, at least,
and probably on the real planet also,
slight submarginality is insufficient to hold the jets
realistically straight because it allows long-wave meandering
to be too easily excited, even if not quite self-excited.
Substantial submarginality with nonmonotonic PV gradients
restricts $\LD$ values to be substantially less than the jet
spacing, as suggested schematically
in Fig.~\ref{fig:cartoon}, at top right.
A similar restriction probably holds on the real planet,
constraining \whl\ depths and
therefore abundances such as that of water.
And the submarginality depends on
having deep jets
(Section~\ref{sec:mechs-migration}\ref{subsec:nodeepjets}
below).

The plan of the paper is as follows.
Section~\ref{sec:model-formulation}
introduces the model.  Section~\ref{sec:forcing}
introduces the PV-biased forcing
and shows how it can act \qf ally.
Section~\ref{sec:parameters} motivates
our choice of parameters ---
emphasizing those choices,
including submarginal $\LD$,
that lead to realistic, quasi-zonal \whl\ structures.
Section~\ref{sec:mainresults}
surveys the main body of results.
Section~\ref{sec:mechs-migration}
discusses the chaotic
vortex-interaction mechanisms
that produce realistic structures.
We find that
migration
is crucial.
By contrast, upscale energy cascades
and the Rhines mechanism play no significant role,
even though the flow is turbulent in the accepted
chaotic-dynamics
sense.
We also point out, in
Section~\ref{sec:mechs-migration}\ref{subsec:noncascade},
that our model's \qf al timescales are considerably
shorter than the
relevant radiative timescales, those near the interface
in Fig.~\ref{fig:cartoon}.

Section~\ref{sec:kelvin} shows that
another mechanism ---
the Kelvin passive-shearing  mechanism \citep{Thomson1887},
much discussed in recent years
under the headings ``CE2'', ``SSST'', and ``zonostrophic
instability''
\citep[e.g.,][\andrefs]{Srinivasan2014} ---
has interesting effects but is
unable to produce
realistic \whl\ structures
in our model.
Section~\ref{sec:conclu}
presents some concluding remarks and suggestions
for future work.

\section{Model formulation}
\label{sec:model-formulation}

We use a doubly-periodic, \qg, pseudo\-spectral
$\beta$-plane version of the
\oneandahalflayer\ model,
with leapfrog timestepping and a weak Robert filter.
The model tries to mimic
conditions in a band of northern-hemispheric latitudes
containing two deep jets,
one prograde and one retrograde.
The simplest way to achieve shear-stability
properties resembling those of a horizontally larger domain
\citep[][\S2.2.1]{mythesis}
is to choose
the model's zonal ($x$) to
latitudinal
($y$) aspect ratio to be 2:1.
In most runs a 512$\times$256 spatial grid is used.
Further detail is in
\citet{mythesis},
and an annotated copy of the code is provided online
through the authors' websites.

Figure~\ref{fig:cartoon}
shows schematically a
meridional
slice through the model, with the
upper or \whl\
jets shown stronger
than the deep jets and the interface correspondingly tilted,
as dictated by thermal-wind balance,
with elevation $\zeta$ say.
The $y$ axis points northward
and the $x$ axis eastward, out of the paper.
The central raised, i.e., cold, interface is in a model belt,
cyclonically sheared, with
model zones on either side and with
the whole structure repeated periodically.
The underlying dry-convective layer is modeled as
adiabatic and infinitely deep, with constant
potential temperature $\theta$.
The constant-$\theta$ interface with the \wl\ is
shown as the bottom of the shaded region.
The~interface is flexible
and responsive to the dynamics.
The shading is meant to indicate
the stable stratification
suggested by the work of \citet{Sugiyama2014}, 
concentrated
near the base of the \wl\ though
less so in zones than in belts.
Belt centre is where
the interface
is highest,
bringing it closest to 
the lifting condensation level
for water.
Such a configuration is consistent with
thermal-wind balance
and with the standard perception
that the \wl's stable stratification --- a
positive vertical gradient
$\partial\sliver\theta\antisliver/\partial z$
--- results from moist
convection with the convection strongest in the belts.

The model equations
for $\qq(x,y,t)$, the \wl's
large-scale \qg\ potential vorticity
(PV),
with forcing $\forcing(x,y,t)$ and
small-scale dissipation $\dissipation(x,y,t)$,
are
\begin{eqnarray}
\left(
   \frac{\partial}{\partial t}
   + 
   \bu\cdot{\nabla}
\right)
\qq
\;=\;
\forcing + \dissipation
\;,
\phantom{ssss}
\label{eq:pv-eq}
\\
\qq
\definedas
\nabla^2 \psi + \beta y
-
\kd^2
\left(
   \psi - \psideep
\right)
.
\label{eq:pv-def}
\end{eqnarray}
Here $\nabla^2$ is the two-dimensional Laplacian
in the $xy$ plane,
$\beta$ is the
local latitudinal gradient of the vertical
component of the planetary vorticity,
$\kd$ is the reciprocal of
the Rossby deformation length $\LD$ based on the
\whl's mean depth and on $g'$, the reduced
gravity at the interface;
$\psi(x,y,t)$ and $\bu(x,y,t)$
are the geostrophic streamfunction and velocity
for the
\wl,
such that $\bu$ is horizontal with components 
$
\bu
= (\uu,\,\vv)
= (-\partial\psi/\partial y,\;\partial\psi/\partial x)
$,
and
$\psideep$ is the
geostrophic streamfunction
for the prescribed steady,
zonally-symmetric zonal flow
$\uudeep = -\partial\psideep/\partial y$
in the dry-convective layer.
\ $\dissipation$ is a quasi-hyperdiffusive dissipation
in the form of a
high-wavenumber spectral filter,
used only to maintain
numerical stability.
It will be ignored in most of the theoretical discussion.
We
adopt the filter described in appendix B of \citet{Smith2002}.
The model code evaluates
$\nabla\psi$ and $\nabla\qq$ in spectral space before
FFT-transforming to physical
space and evaluating $\bu\cdot{\nabla}\qq$ by
pointwise multiplication,
then transforming back.

Following \citet{Dowling1993} and \citet{Stamp1993},
we somewhat arbitrarily take the deep flow
to have
a sinusoidal profile
\begin{equation}
\uudeep(y)
\;=\;
U_0
\,+\,
\Umax \sliver \sin
\left(
   \frac{y}{L}
\right)
\,,
\label{eq:deep-jets}
\end{equation}
where $U_0$ and $\Umax$
are constants.
The lengthscale
$L$ is  $(2\pi)^{-1}$ times the
domain's $y$-period, the full wavelength of the jet spacing,
which we fix at 10,000\km\ to represent mid-latitude conditions.

The real deep-jet profiles may of course be different.
However, they are not well known.
DI's
analysis did, to be sure,
find rounded
$\uudeep(y)$ profiles, in striking contrast with the
sharper profiles
found in some dry-convective models.
However, DI's cloud-wind analysis may not have been accurate enough
to fix $\uudeep(y)$
with great precision.  While cloud-wind analyses have
greatly improved since then
--- see especially \citet{Asay-Davis2009} ---
we are not aware of any corresponding published estimates of
$\uudeep(y)$ profiles and their error bars.

With the exception of $\qq$, which contains
the non-periodic terms
$\beta y - \kd^2U_0y$,
all the model's \whl\ fields
are assumed to be
doubly-periodic
including
the streamfunction $\psi$ and the
zonal-mean gradient $\partial\qqbar/\partial y$
of $\qq$,
\begin{equation}
\frac{\partial\qqbar}{\partial y}
\;=\;
\beta
\,-\,
\frac{\partial^2\uubar}{\partial y^2}
\,+\,
\kd^2
\left(
   \uubar - \uudeep
\right)
\;.
\label{eq:mean-pv-grad}
\end{equation}
The periodicity of $\psi$
entails that
\begin{equation}
\int_0^{2\pi L}\Antisliver\uubar\Sliver dy = 0
\;,
\label{eq:ubar-int-zero}
\end{equation}
which implicitly assumes not only that we are in a
particular reference frame, but also that the
domain-averaged angular momentum budget is
steady.\footnote{In
other words, any domain-averaged
external zonal force is either negligible or balanced by
a domain-averaged ageostrophic mean $y$-velocity.
  }
And
without loss of generality we may take the domain integrals
of $\psi$ and $\forcing$ to
vanish:
\begin{equation}
\iint\antisliver\psi\Sliver dxdy = 0
\,,
\quad
\iint\antisliver\forcing\Sliver dxdy = 0
\,.
\label{eq:domain-ints-zero}
\end{equation}
The
first of these follows from the freedom to
add an arbitrary function of time $t$ alone to the
streamfunction $\psi$, with no effect on the \qg\ dynamics.
Physically, this says that a small
variation in the total mass of the \wl,
due for instance to horizontally-uniform
diabatic processes,
has no dynamical effect as long as the
mean \whl\ depth,
hence $\LD$ value, can be considered constant
to leading order in Rossby number.
From (\ref{eq:pv-def}) and the first of (\ref{eq:domain-ints-zero})
we then have
\begin{equation}
\frac{\partial }{\partial t}
\iint\antisliver\qq\Sliver dxdy
\;=\; 0
\;,
\label{eq:pv-impermeability}
\end{equation}
which is consistent with (\ref{eq:pv-eq}) only if
the second of
(\ref{eq:domain-ints-zero}) also holds.
This can be seen
by domain-integrating
the flux form of (\ref{eq:pv-eq})
and noting
that
$\vvbar = \partial\psibar/\partial x = 0$, and that
$\iint\Antisliver\dissipation\sliver dxdy = 0$
in virtue of $\dissipation$'s restriction to the highest wavenumbers.
It is convenient to view (\ref{eq:pv-impermeability}) as
a \qg\ counterpart to the ``impermeability theorem'' for the
exact, Rossby--Ertel PV \citep[e.g.,][]{Haynes1990}.

From here on we ignore the small-scale dissipation $\dissipation$.
The zonal-mean dynamics is then described by
\begin{equation}
\frac{\partial \qqbar}{\partial t}
\;=\;
-\Sliver
\frac{\partial(\overline{\vv'\qq'})}{\partial y}
\;+\;
\forcingbar
\;,
\label{eq:mean-pv-equation}
\end{equation}
where the primes denote departures from zonal averages
$\overline{(\phantom{n})}$.
The model's Taylor identity
\citep[e.g.,][]{Buhler2014b},
which allows the mean PV dynamics to be translated into
mean momentum dynamics, if desired,\footnote{The
mean momentum dynamics is given by the
minus the indefinite $y$-integral of (\ref{eq:mean-pv-equation}),
in which
$
\int \Antisliver dy \sliver
(
  \kd^2 \partial\psibar\antisliver/\partial t
  +
  \forcingbar
)
$
represents minus
the Coriolis force from the
ageostrophic mean $y$-velocity,
whose $y$-derivative is related via mass conservation to
layer-depth changes and to the forcing $\forcingbar$ conceived of as
mass injection or withdrawal.
  }
is
\begin{equation}
\Sliver
\frac{\partial(\overline{\uu'\vv'})}{\partial y}
\;=\;
-\,
\overline{\vv'\qq'}
\;.
\label{eq:taylor-identity}
\end{equation}
The Taylor identity is
a consequence of
(\ref{eq:pv-def}) alone, as is
easily verified
using
$\partial\overline{(\phantom{n})}/\partial x = 0$,
hence valid at all eddy amplitudes and
independent of forcing and dissipation.
By multiplying (\ref{eq:pv-eq}) by  $-\psi$,
and continuing to ignore $\dissipation$,
we find
the relevant
energy equation
to be
\begin{equation}
\frac{\partial E}{\partial t}
\;=\;
-\Antisliver
\iint \Antisliver\psi\sliver\forcing\Sliver dxdy
\label{energy_equation}
\end{equation}
where
$E$ is the kinetic plus available potential energy of the \wl\
--- the model's only variable energy ---
divided by the mass per unit area.
That is,
\begin{equation}
E
\definedas
\iint 
\left (
   \half  |\nabla\psi|^2
   +
   \half
   \kd^2 \psi^2
\right )
dxdy
\;.
\label{eq:energy-def}
\end{equation}

\section{The forcing $\forcing$}
\label{sec:forcing}

\subsection{Impulsive injection of small vortices}
\label{sec:injection}

The forcing $\forcing$ corresponds to
repeated injections of
close-spaced,
east-west-oriented pairs of small
vortices at random locations
and in alternating order,
cyclone-anticy\-clone
alternating with
anticyclone-cyclone.  In each pair,
the cyclone is weaker than the anticyclone by a fractional
amount $\fractionalbias$, say, which
we call the ``fractional bias'', and which increases with
vortex strength so as to
express the notion that
the dry-convective layer
supplies the \wl\ with
mass and heat but with relatively more mass in the
stronger convection events.
Each vortex is impulsively injected
using a parabolic PV profile,
Eq.~(\ref{eq:injection-parabola})
and inset to
Fig.~\ref{fig:injections-before-saturation} below.

We acknowledge that this
must be an exceedingly crude representation
of vortex generation by
the real three-di\-men\-sion\-al
convection, whose structure and
tangled vortex-line topology
are unknown and have yet to be
plausibly modeled.
A simplistic vortex-injection
scheme may be the best that can be done within the
\oneandahalflayer\ dynamics,
and indeed is a rather time-honored idea 
(e.g.,
\citeauthor{Vallis1997}
\citeyear{Vallis1997}, Section~3b;
\citeauthor{Li2006}
\citeyear{Li2006};
\citeauthor{Showman2007}
\citeyear{Showman2007};
\citeauthor{Humphreys2007}
\citeyear{Humphreys2007},
Section~5a and Fig.~6).
Within such a scheme it is
arguably most realistic to use vortices of both signs,
avoiding the
anticyclones-only scenarios that might be suggested by too
exclusive a focus on cloud-top observations and which,
in any case,
would correspond to mass injection only.

We use east-west-oriented pairs
for two reasons.  One is
to make zonally
averaged quantities such as $\qqbar$ and $\overline{\vv'\qq'}$
less noisy.  The other, less obvious, reason is
an interest in assessing whether the
Kelvin passive-sheared-disturbance
mechanism (also discussed under the headings
``CE2'' and ``SSST'' in the literature)
has a significant role in any of the regimes we find.
The Kelvin mechanism operates when
the
in\-ject\-ed vortices are so
weak that they are
passively sheared by the mean flow $\uubar(y)$,
producing systematically slanted structures
and hence
Reynolds stresses
$\overline{\uu'\vv'}$
potentially able to
cause jet self-sharpening by ``zonostrophic instability''.

The Kelvin mechanism is
entirely different from the
inhomogeneous-PV-mixing
mechanism that produces
terrestrial strong jets,
through drastic piecewise rearrangement of a background PV gradient.
It is also different from
the Rhines mechanism, in which the injected vortices are
strong enough to undergo the usual vortex interactions,
especially the vortex merging
that produces an
upscale energy cascade that is then arrested, or
slowed, by the Rossby-wave elasticity of an
\emph{un-rearranged},
uniform
background PV gradient.
We are of course interested in whether
any of these mechanisms have significant roles.
Regarding the Kelvin mechanism,
it is strongest when the forcing is anisotropic in the sense of
east-west vortex-pair orientation
\citep[e.g.,][]{Shepherd1985, Srinivasan2014}.
So our choice of
east-west orientation will give the Kelvin mechanism its best chance.

The impulsive vortex injections,
corresponding theoretically to temporal delta functions in
the forcing function $\forcing$, are actually spread
over time intervals $2\Deltat$ to avoid exciting the
leapfrog computational mode,
where $\Deltat$ is the timestep.
This is still fast enough
for advection to be negligible, implying that the injections are
instantaneous to good approximation.  The parabolic
profile of the resulting change 
$\Delta\qq(\rr)$ in the PV field is given by
\begin{equation}
\Delta\qq(\rr)
\;=\;
\qqpeak
\left(
   1 - \frac{\rr^2}{\rr_0^2}
\right)
\hspace{0.4cm}
(\rr \leqslant \rr_0)
\;,
\label{eq:injection-parabola}
\end{equation}
the
peak vortex strength
$\qqpeak$ being positive for a cyclone
and negative for
its accompanying
anticyclone.  The relative radius
$\rr := |\bx - \bxc|$ with
 $\bx=(x,y)$ denoting horizontal position and
$\bxc=(\xc,\yc)$ the position of the
vortex center.
The radius $\rr_0$ is taken as small as we dare, consistent with
reasonable resolution and realistic-looking vortex interactions
\citep[][\S\S2.1.1, 3.5.2]{mythesis}.
In most cases
$\rr_0=4\Deltax$
where $\Deltax$ is the grid size.

\subsection{The complementary forcing}
\label{subsec:complementary-forcing}

Thanks to the peculiarities
of \qg\ dynamics and to our model choices
we need the forcing to satisfy (\ref{eq:domain-ints-zero}b).
The model code does this automatically,
by assigning zero values to all spectral
components having total wavenumber zero.
The most convenient way to
see what it means, however, is to think of each
injected vortex as satisfying (\ref{eq:domain-ints-zero}b)
individually.  When, for instance,
a small anticyclone is injected, it is accompanied by
a domain-wide cyclonic ``complementary forcing'',
in the form of a small, spatially-constant contribution,
additional to (\ref{eq:injection-parabola})
and spread over the entire domain, such that
$\iint\antisliver\forcing\Sliver dxdy$
is zero.
That is not to say that the \emph{dynamical} response
to a single injection is domain-wide.
Rather,
the complementary forcing
is
no more than a convenient bookkeeping
device to guarantee that the forcing is consistent,
at all times,
with our choice of model setup including
the choice (\ref{eq:domain-ints-zero}) and its consequence
(\ref{eq:pv-impermeability}).

Consider for instance a localized mass injection. The dynamical
response is formation of an anticyclone, namely a negative
anomaly in the $\qq$ field together with the associated
mass and velocity
fields obtainable by PV inversion
\citep[e.g.,][\andrefs]{Hoskins1985}.
Those fields describe an outward
mass shift and anticyclonically-circulating winds,
the whole structure extending outwards and
decaying exponentially on the lengthscale $\LD$.
The complementary forcing is
quite different.  It can be pictured
as a uniform, domain-wide
withdrawal of a compensating amount of mass
that is
small, of the order of the Rossby number,
and
has no dynamical effect whatever.
The complementary forcing
is an artificial device
to keep the mass of the model \wl\ exactly constant.
Of course with anticyclonic bias
a domain-wide mass withdrawal would have to occur in
reality, presuming statistical steadiness,
and would involve
cloud physics and
radiative heat transfer
\citep[e.g.,][]{Li2006}.
However, that aspect
of the problem is invisible to the
\qg\ dynamics.\footnote{A
reviewer's comment prompts us to remark that
spatially uniform layer-mass changes
are invisible
to
\qg\ \emph{channel} dynamics also.
The first of (\ref{eq:domain-ints-zero}), and the remarks below
(\ref{eq:domain-ints-zero}), still apply.
  }

\subsection{The vortex-injection scheme}
\label{subsection:injection}

We have explored many vortex-injection schemes, with
many choices of the way in which
injected vortex strengths
$|\sliver\qqpeak|$ and fractional bias $\fractionalbias$
are made to increase
with the interface elevation or coldness $\zeta$.
The simplest choices, with strengths increasing monotonically
with $\zeta$, produced
runaway situations with
vortices far stronger than the
real planet's mean shears and observed vortices,
incompatible with our aim of
finding flow regimes that are both
realistic and statistically steady.

After much experimentation,
the following vortex-injec\-tion scheme proved successful,
one aspect
of which is that
$|\sliver\qqpeak|$
is never allowed to exceed a set value
$\qqmax>0$.
The sensitivity to 
interface elevation or coldness is set by
a parameter
$\psilim>0$, in terms of which
the definition of $\zeta$ will be written as
\begin{equation}
\zeta(x,y,t)
\;=\;
\frac{\psideeptilde(y)-\psi(x,y,t)}{\psilim}
\,,
\label{eq:def-psi-lim}
\end{equation}
where
\begin{equation}
\psideeptilde(y)
\;\definedas\;
L\sliver\Umax\sliver\cos
\left(
   \frac{y}{L}
\right)
\;,
\label{eq:def-psideeptilde}
\end{equation}
corresponding to the $y$-oscillatory or jet-like part of the
deep flow (\ref{eq:deep-jets}).\footnote{Because
of its double periodicity, our idealized model has
no way of representing large-scale gradients in
$\psideep$ except insofar as
$d\sliver\psideep/dy=-\uudeep$
enters the background PV
gradient
(\ref{eq:mean-pv-grad}).
Of course
the model also ignores the real planet's other large-scale gradients,
and \mbox{associated} mean meridional circulations,
for instance large-scale gradients
in temperature, in composition
including hydrogen ortho--para fraction
\citep[e.g.,][Fig.\;10]{Read2006}, and in the Coriolis parameter
and $\LD$.
  }
Injections are
done one pair at a time, with the intervening
time intervals selected at random from a specified range
$[4\Deltat,\,\ttmax]$,
with uniform probability.
The minimum value $4\Deltat$
ensures that injection events
do not overlap in time.
The maximum value $\ttmax$
is usually chosen to be much larger, such that
$\half\ttmax$, close to the average time interval, is
of the same order as the
background shearing time
$L/\Umax$.
We interpret these temporally sparse injections as idealizing
the intermittency of real convection, probably governed by
slow but chaotic dry-convective dynamics along with
time-variable structure near the interface
\citep[e.g.,][]{Showman2005, Sugiyama2014}.

\begin{figure} [t]
\includegraphics[width=21.5pc,angle=0]
{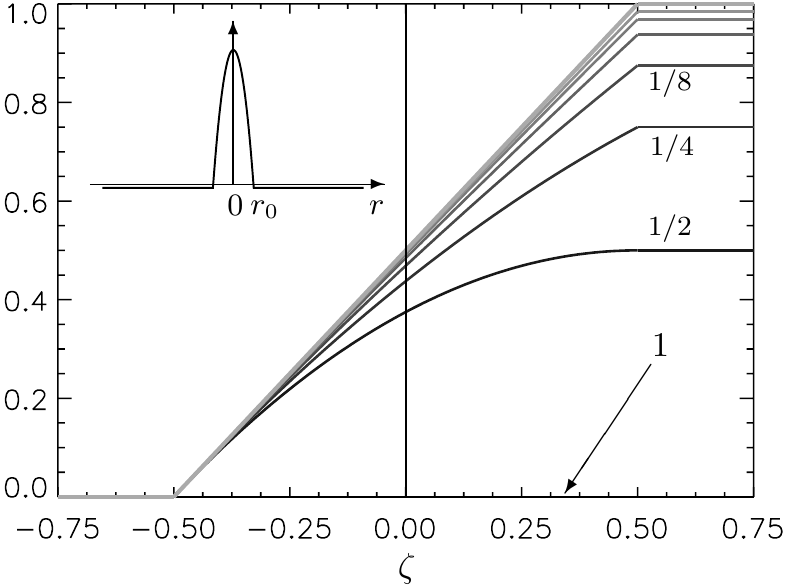}
  \vspace{-0.6cm}
\caption{\footnotesize
Functions used in the
vortex-injection scheme.  The inset at top left shows the
$\Delta\qq(\rr)$ profile for an injected cyclone of radius
$\rr_0 \ll \LD$,
discussed in Eq.~(\ref{eq:injection-parabola})ff.
The top, light-colored curve in the main figure shows
the function
$
\ramp(\zeta)
$
defined
in (\ref{eq:ramp}),
equivalently $|\sliver\qqpeaka|/\qqmax$
when the choice (\ref{eq:anticyclones-before-saturation})
is made.
The remaining curves show
$\sliver\qqpeakc/\qqmax$ for different nonzero
biases, according to
(\ref{eq:bias-defs})--(\ref{eq:anticyclones-before-saturation})
and the text below (3.6).
From bottom up we have
$\fractionalbias=$ 1 ($\qqpeakc=0$
for all $\zeta$) and then $\qqpeakc/\qqmax$ for
$\fractionalbiasmax=$ 1/2, 1/4, 1/8, 1/16, 1/32, 1/64.
Note, however, that (\ref{eq:anticyclones-before-saturation}) is used
subject to the further conditions described in
(\ref{eq:saturation-constraint})ff.
        }
\label{fig:injections-before-saturation}
\end{figure}

For each injection event
a location $\bx=(x,\,y)$ is chosen at random
and a close-spaced but non-overlapping pair of vortices,
each of radius $\rr_0$ as specified in
(\ref{eq:injection-parabola}),
is injected at the pair of neighboring points
\begin{equation}
(\xc,\,\yc)
\;=\;
\mbox{$(x \pm \half\s,\; y)$}
\;.
\label{eq:injection-points}
\end{equation}
where the separation $\s$ is fixed at
\begin{equation}
\s
\;=\;
2\rr_0 + \Deltax
\;.
\label{eq:injection-separation}
\end{equation}
We denote the respective strengths by
$\qqpeak=\qqpeakc>0$ for the cyclone and
$\qqpeak=\qqpeaka<0$ for the anticyclone,
with magnitudes
always
in the ratio
$\oneminusbias\leqslant 1$ where
\begin{equation}
\oneminusbias
\;=\;
\left|
   \frac{\qqpeakc}{\qqpeaka}
\right|
\;=\;
(1 - \fractionalbias)
\;.
\label{eq:bias-defs}
\end{equation}
The fractional bias $\fractionalbias$
is either 1, to give anticyclones only,
as in \citet{Li2006} and in \citet{Showman2007}, or
\begin{equation}
\fractionalbias
\;=\;
\fractionalbias(\zeta)
\;=\;
\fractionalbiasmax\Sliver\ramp(\zeta)
\label{eq:fracbias}
\end{equation}
in all other cases,
where $\fractionalbiasmax<1$ is a positive constant
and where $\ramp(\zeta)$ is the
three-piece ramp function defined by
\begin{equation}
\ramp(\zeta)
\definedas
\left\{\hspace{-5pt}
  \begin{array}{cc}
         0            &~~~ \mbox{ $(\zeta\leqslant -\half)$},\\
   \half \,+\, \zeta  &~~~ \mbox{ $(-\half\leqslant\zeta\leqslant\half)$},\\
         1            &~~~ \mbox{ $(\zeta\geqslant\half)$}
\;.
   \end{array}
\right.
\label{eq:ramp}
\end{equation}
This function is plotted as the top, light-colored curve in the main part of
Fig.~\ref{fig:injections-before-saturation}.

It remains to choose how $|\sliver\qqpeaka|$
varies.
The simplest choice would be
\begin{equation}
|\sliver\qqpeaka|
\;=\;
\qqmax\Sliver\ramp(\zeta)
\;,
\label{eq:anticyclones-before-saturation}
\end{equation}
so that $|\sliver\qqpeaka|/\qqmax$ is given by the
light-colored curve
in Fig. 2.  Then $\sliver\qqpeakc/\qqmax$ is given by the
curves underneath, for
values of $\fractionalbiasmax < 1$, plotted using
(\ref{eq:anticyclones-before-saturation}) with
(\ref{eq:bias-defs})--(\ref{eq:ramp}).
The label 1 marks the null curve
$\sliver\qqpeakc/\qqmax=0$ for
the anticyclones-only
case $\fractionalbias=1$.
However, the choice
(\ref{eq:anticyclones-before-saturation})
still produces runaway situations
incompatible with statistical steadiness,
except when $\qqmax$ is
made too small to produce
significant small-scale vortex activity.
For larger $\qqmax$ values,
enough to produce
such activity,
the typical behavior is
the growth and unbounded
strengthening of a large cyclone.
The large cyclone's cold-interface footprint
in $\zeta$,
still larger in area,
induces strong local injections from which the
small injected
cyclones tend
to migrate inward and the small
anticyclones outward
to give a cumulative, and apparently unbounded,
increase in the
large
cyclone's size and strength.
(Notice by the way that this mechanism is quite different from
the classic vortex merging or upscale energy
cascade.
The possibility of an unbounded
increase
in vortex strength is
another peculiarity
of \qg\ theory,
predicting its own breakdown as Rossby numbers increase.)

Large, strong cyclones have correspondingly large
$\qq'$ values,
motivating our final choice,
which is to use
(\ref{eq:bias-defs})--(\ref{eq:anticyclones-before-saturation})
--- remembering that
$\qqpeakc = \oneminusbias\sliver|\sliver\qqpeaka|$ ---
whenever the local $\qq'$ value satisfies
\begin{equation}
\max
\big(
   \oneminusbias\sliver|\sliver\qqpeaka| + \qq', \
   |\sliver\qqpeaka| - \qq'
\big)
\;\leqslant\;
\qqmax
\label{eq:saturation-constraint}
\end{equation}
whereas, if
(\ref{eq:anticyclones-before-saturation})
gives a $|\sliver\qqpeaka|$ value
that makes the left-hand side
of (\ref{eq:saturation-constraint})
greater than $\qqmax$,
then $|\sliver\qqpeaka|$ is reduced just enough
to achieve equality, i.e., reduced just enough to satisfy
(\ref{eq:saturation-constraint})
with
$\oneminusbias$
unchanged.
The second argument of the $\max$ function
in (\ref{eq:saturation-constraint})
covers the possibility
that strong anticyclones with large negative $\qq'$ might
occur, though it is the first argument that
prevails in all the cases we have seen.

The limitations thus placed on
the strongest vortices injected
are interpreted here as
reflecting not only the limitations of \qg\ theory,
but also the unknown limitations of
the real, three-dimensional moist convection
as a mechanism for generating
coherent vortices
on the larger scales represented by our model.
On smaller scales one must expect
three-dimensionally turbulent vorticity fields
with still stronger peak magnitudes --- as terrestrial tornadoes
remind us --- though, with no solid lower surface,
the details are bound to be different.
For one thing, net mass injection rates are
bound to be modified by such phenomena as
evaporation-cooled,
precipitation-weighted
thunderstorm down\-drafts,
also called
microbursts,
contri\-but\-ing
negatively.
The concluding remarks
in Section~\ref{sec:conclu}
will suggest a possible
way of replacing
(\ref{eq:saturation-constraint})
by something less artificial,
albeit
paid for by
further expanding the model's parameter space.

\subsection{\Qf al effects}
\label{subsec:qf}

As mentioned earlier, the bias $\fractionalbias$ has
\qf al\ effects.  These are most obvious
in the zonal-mean dynamics
described by (\ref{eq:mean-pv-equation}),
under the constraints
(\ref{eq:ubar-int-zero})--(\ref{eq:pv-impermeability}).
Because of thermal-wind balance and the
positive slope of the ramp function $\ramp(\zeta)$,
the sign of $\forcingbar(y,t)$ tends
on average
to be anticyclonic in belts and cyclonic in
zones whenever the
upper or \whl\
jets are stronger than the deep jets
(\ref{eq:def-psideeptilde}),
the case sketched
in Fig.~\ref{fig:cartoon}.
The converse holds in the opposite case.  So $\forcingbar$ tends
on average
to reduce
differences between shears in
the upper jets
and in the deep jets.
There is a corresponding \qf al effect on
large cyclones.
By contrast,
fluctuations such as those giving rise to
the eddy-flux term 
$-\partial(\overline{\vv'\qq'})/\partial y$
in
(\ref{eq:mean-pv-equation})
can act
in the opposite sense,
in some cases giving rise to
realistic interface-temperature structures
in the manner sketched in Fig.~\ref{fig:cartoon}.

We find that the
\qf al\ effects can be understood alternatively
from environment-dependent
negative contributions to the
right hand side of the energy equation
(\ref{energy_equation}), competing with the positive,
environment-in\-de\-pen\-dent
``self-energy'' inherent in each injection.
This
contrasts with the standard, perfectly unbiased
forcing used in beta-tur\-bu\-lence theory
\citep[e.g.,][]{Srinivasan2012a},
which is designed such that the
self-energy is the only contribution,
allowing one to prescribe
a fixed, positive
energy input rate
$\varepsilon$,
which along with spectral narrowness
is the normal prelude to using
Kolmogorovian arguments.
However,
it would then be necessary to introduce
a separate large-scale dissipation
term, as would be necessary also
if cyclonic bias, $\fractionalbias < 0$,
were to be used in our scheme.
(Not surprisingly,
taking $\fractionalbias < 0$
has antifrictional effects.
When we tried it, the most conspicuous result was
self-excitation of unrealistic long-wave undulations.)

When $\forcingbar$ and other \qf al effects
with $\fractionalbias > 0$
are strong enough to produce
realistic,
statistically steady flow regimes,
we find that upper-jet
profiles tend to be pulled fairly
close to deep-jet profiles.
This tendency shows up robustly in test
runs initialized with upper jets
both weaker and stronger than the deep
jets.  In most cases, therefore, we
use a standard initialization
in which
the upper-jet profiles are the same as
the deep-jet profiles
(\ref{eq:def-psideeptilde}),
making $\zeta=0$ to start with, and
the average forcing spatially uniform.
We then observe how the upper profiles,
$\zeta$,
and the implied forcing all
change in response to
the eddy flux $\overline{\vv'\qq'}$
in (\ref{eq:mean-pv-equation}).

\section{Parameter choices}
\label{sec:parameters}

\subsection{Sensitivity}
\label{subsec:sensitivity}

It turns out that the interesting cases,
statistically stea\-dy
with realistic, quasi-zonal
$\zeta$ or interface-temperature structure,
occupy only a small region within
the model's vast parameter space.
Not surprisingly, the behavior is sensitive to
$\qqmax$ and $\LD$ values, which govern
the strength and nature of
the model's vortex activity
all the way from cases with no such activity ---
having only the
Kelvin (CE2/SSST) passive-shearing mechanism ---
up to cases with vortex activity so violent as to disrupt the
quasi-zonal $\zeta$  structure
altogether.  It turns out that
the Kelvin mechanism
is unable to produce realistic $\zeta$ structure.
See Section~\ref{sec:kelvin} below.
The most interesting cases, our main focus,
turn out to be
those exhibiting
chaotic vortex interactions
just strong enough to
make an impact on the $y$-profiles
of $\overline{\vv'\qq'}$
in (\ref{eq:mean-pv-equation}).

A big surprise, though,
was that the behavior is very sensitive to the
choice of $\fractionalbiasmax$, with the most interesting cases
clustered around small values $\lesssim10^{-1}$.
This was especially surprising in view of the past work of
\citet{Li2006} and \citet{Showman2007}
using purely anticyclonic forcing
$\fractionalbias = 1$.
The different behavior seems related
in part to the absence of deep jets in their studies but
presence in ours, in combination with
the \qf al $\forcingbar$ effect.  Some further
discussion is given in
Section~\ref{sec:mechs-migration}\ref{subsec:nodeepjets}
below.

\subsection{$\LD$ values and A2 stability}
\label{subsec:ld-a2}

In considering choices of $\LD$, and remembering its
latitude dependence,
we would like to respect observational as well as theoretical
constraints.
However,
observational constraints from the comet-impact
waves are controversial and unclear.\footnote{One
reason is that
even if the comet-impact waves were gravity waves
guided by the \wl\, they would have had
a different structure in the underlying dry-convective layer,
more like surface-gravity-wave
structure than that of \oneandahalflayer\ dynamics.
Another reason is the case made by \citet{Walterscheid2000}
that the observed comet-impact waves were in any case
more concentrated in Jupiter's stratosphere.
  }
Also, observational constraints
from the DI work and its successors
apply mainly to the lower latitudes of the Great Red Spot and
other large anticyclonic Ovals, roughly 15\degree--35\degree.
The original DI work
suggested
$\LD$ values at, say, 35\degree S,
that were
not strongly constrained but were
estimated as roughly in the range 1500--2250\km.
The more recent work of \citet{Shetty2010},
based on a much more sophisticated cloud-wind methodology,
appears to constrain $\LD$ values more tightly,
for instance producing values close to 1900\km\
at latitudes around 33.5\degree S
from an analysis of the flow around a large anticyclone,
Oval BA.
However, unlike DI, who used
\oneandahalflayer\ primitive-equation dynamics,
\citet{Shetty2010} assumed that \qg\
\oneandahalflayer\ dynamics
applies accurately to the real planet.
In any case, it is likely that all these estimates apply to
the locally deeper \wl\ expected near large anticyclones,
suggesting somewhat smaller $\LD$ values
further eastward or westward, as well as further poleward
in virtue of the increasing Coriolis parameter.

Our approach will be to reserve judgement on
these issues, and simply to
find a range of  $\LD$ values for which
the idealized model behavior looks realistic.

As indicated near the end of Section~\ref{sec:intro},
we need to keep the
model's jets straight by excluding long-wave shear instability.
With nonmonotonic upper PV gradients
our model has a shear-instability threshold
strongly influenced by the value of $\LD$, as does,
almost certainly, the real planet as well.
Linear theory shows that, when the threshold
is slightly exceeded, the instability first kicks in as
a long-wave undulation, phase-coherent between adjacent jets,
a fact that we have cross-checked
in test runs with the unforced
model showing, in addition,
that the undulation equilibrates nonlinearly
to a moderately small amplitude
\citep[][\S4.1]{mythesis}
without otherwise disturbing the PV distribution, to any
significant extent.
Such a phase-coherent long-wave undulation would be
conspicuous on the real planet but is not observed.
To get straight jets we must stay below the
shear-instability threshold.

In our model, for upper-jet profiles
kept close to the deep-jet profiles
(\ref{eq:def-psideeptilde})
by the \qf al $\forcingbar$ effect,
the upper PV gradients
are indeed nonmonotonic, and strongly so
if we take plausible values of $\Umax$
and $\beta - \kd^2U_0$.
Recall Eqs.~(\ref{eq:deep-jets})--(\ref{eq:mean-pv-grad}),
noting that
the $y$-oscillatory part of
the $\kd^2$ term in (\ref{eq:mean-pv-grad})
is small when the upper and lower jet
profiles are close,
that is, when
$\uubar(y)$ is close to $\uudeep(y) - U_0$.

If for instance we take
$\Umax \simeq 30$\ms\ and
$\beta - \kd^2U_0$
anywhere between the value
zero suggested by DI's results and
the value of $\beta$ itself at the equator,
$\simeq5\times 10^{-12}$\smmone,
then we get strongly nonmonotonic
$\partial\qqbar/\partial y$
essentially because, with
$L=(2\pi)^{-1}\times10,000$\km\ = 1592\km,
we have
$\partial^2\uubar/\partial y^2$ values in the
range $\pm\Umax/L^2=\pm12\times 10^{-12}$\smmone,
whose magnitude is well in excess of $\beta$ at any latitude.

The model's jet flow is then shear-unstable for sufficiently large
$\LD/L$ but stabilized when
$\LD/L$ is taken below the threshold already mentioned,
despite the
strongly nonmonotonic $\partial\qqbar/\partial y$.
The existence of that threshold
was recognized by \citet{Ingersoll1981a} and,
as pointed out by \citet{Dowling1993},
is related to the ``A2 stabilization''
described by Arnol'd's second stability theorem.
It arises because reducing $\LD/L$
reduces the intrinsic phase speeds
and lateral reach
of even the longest, hence fastest possible,
pair of counterpropagating Rossby waves,
each wave propagating upstream on
adjacent prograde and retrograde jets.
These reductions suppress the instability by
``destroying the ability of the two Rossby waves to keep in step''
(\citeauthor{McIntyre1987}
\citeyear{McIntyre1987}, p.\;543; see also
\citeauthor{Hoskins1985}
\citeyear{Hoskins1985}, Fig.~18ff., \andrefs) ---
i.e., to phase-lock,
with each wave holding its own against the mean flow.
That is why the
first wavelength to go unstable
for slightly supermarginal
$\LD/L$ is the \emph{longest} available wavelength, with
zonal wavenumber $\kkmin$, say.

For unbounded or doubly-periodic domains
the A2 theorem says that
the flow is shear-stable
if a constant $\cc$ can be found such that
\begin{equation}
\kd^2 \,+\,\kkmin^2
\;>\:
\frac{\partial\qqbar/\partial y}{\uubar - \cc}
\;,
\label{eq:a2-stability}
\end{equation}
where as before \,$\kd^2\definedas \LD^{-2}$.
For the sinusoidal profiles of our standard initialization,
and for $\beta - \kd^2U_0=0$
as
suggested by DI,
it happens that (\ref{eq:a2-stability}), with $\cc=0$,
is a
necessary as well as a sufficient
condition for stability
\citep{Stamp1993}.
The right-hand side of
(\ref{eq:a2-stability}) is then just $L^{-2}$,
independent of $\Umax$,
and the threshold is precisely at $\kd^2 = L^{-2} - \kkmin^2$.

In our model,
with its 2:1 aspect ratio, we have $\kkmin^2 = L^{-2}/4$.
Therefore,
$\LD\lesssim (4/3)^{1/2}L = 1838$\km\
should
be enough to exclude long-wave shear instability,
as long as the upper profiles $\uubar(y,t)$ stay close to
the deep profiles $\uudeep(y)$.
It is arguable, however, that since the much larger domain
of the real planet should correspond to
$\kkmin^2 \ll L^{-2}$, it might be more appropriate to
take $\LD\lesssim L =  1592$\km.
Having regard to these considerations, we decided to use
$\LD$ values 1500\km\ or less in most of our model runs.
In any case it turned out that
for larger $\LD$ the typical result was
unrealistically strong, or even violent,
long-wave disturbances.

In
sections \ref{sec:mainresults} and \ref{sec:mechs-migration}
we describe and illustrate the model's
behavior for $\LD=$ 1200\km\ and 1500\km\
and for forcings just
strong enough to produce chaotic vortex interactions.
In such cases the model robustly approaches
stable, statistically steady states with fairly straight jets,
and realistic $\zeta$ structures,
over significant ranges
of $\qqmax$ and $\fractionalbiasmax$
and with
nonmonotonic upper PV gradients $\partial\qqbar/\partial y$.

\subsection{Other parameters including $\beta_0$ and $\qqstarmax$}
\label{subsec:pureDI}

We fix $\Umax= 35$\ms\ as a
compromise between low and midlatitude values, and
choose
two values of
\begin{equation}
\beta_0
\definedas
\beta - \kd^2U_0
\;,
\label{eq:def-betazero}
\end{equation}
namely zero and 
$4.03\times 10^{-12}$\smmone.
Both choices
make $\partial\qqbar/\partial y$ strongly nonmonotonic.
The value zero requires prograde $U_0$,
roughly
consistent with DI's results;
see their Fig.~4b
and its idealization in \citet{Stamp1993}.
Prograde $U_0$ is in any case expected
in latitudes
outside a tangent cylinder
of the dry-convective layer
\citep[e.g.,][]{Jones2009}.
The value $4.03\times 10^{-12}$\smmone\ is the value of $\beta$
itself at latitude 35\degree.
For convenience we refer to
these
two cases $\beta_0 = 0$ and
$\beta_0 = 4.03\times 10^{-12}$\smmone\
as ``pure-DI'' and ``midlatitude'' respectively,
remembering, however, that
no\-thing is known about actual $U_0$ values at the higher
latitudes.

Because the \qf al $\forcingbar$ effect
tends to pull our model's
upper jets more or less
close to its deep jets,
the strongest upper mean shears
$\partial\uubar/\partial y$
tend to have
orders of
magnitude
similar to that of
the strongest deep shear $\Umax/L=2.199\times10^{-5}$\smone.
So a convenient
dimensionless measure of $\qqmax$ is
\begin{equation}
\qqstarmax
\,\definedas\,
\frac{\qqmax}{\Umax/L}
~ > \; 0
\;.
\label{eq:qqstarmax-def}
\end{equation}
The parameter
$\qqstarmax$
governs the likely
fate of vortices injected into
background shear of order
$\Umax/L$.
We take values ranging from
$\qqstarmax=0.5$
up to $\qqstarmax=32$.
At the low end of the range,
practically all the injected vortices are shredded,
i.e., sheared passively and destroyed.
(There is still, of course, a
\qf al $\forcingbar$ effect.)
In the highest part of the range, say
$16\lesssim\qqstarmax\lesssim32$,
the strongest injected
vortices all survive even in adverse
shear, e.g., anticyclones in cyclonic shear.
For intermediate values
one typically sees
survival in favorable shear only.
To distinguish the three types of
behavior
we call the
injections
``weak'', ``strong'', and ``semi-strong'' respectively.

\begin{figure} [t]
\hspace{-0.50cm}
\includegraphics[width=22pc,angle=0]
{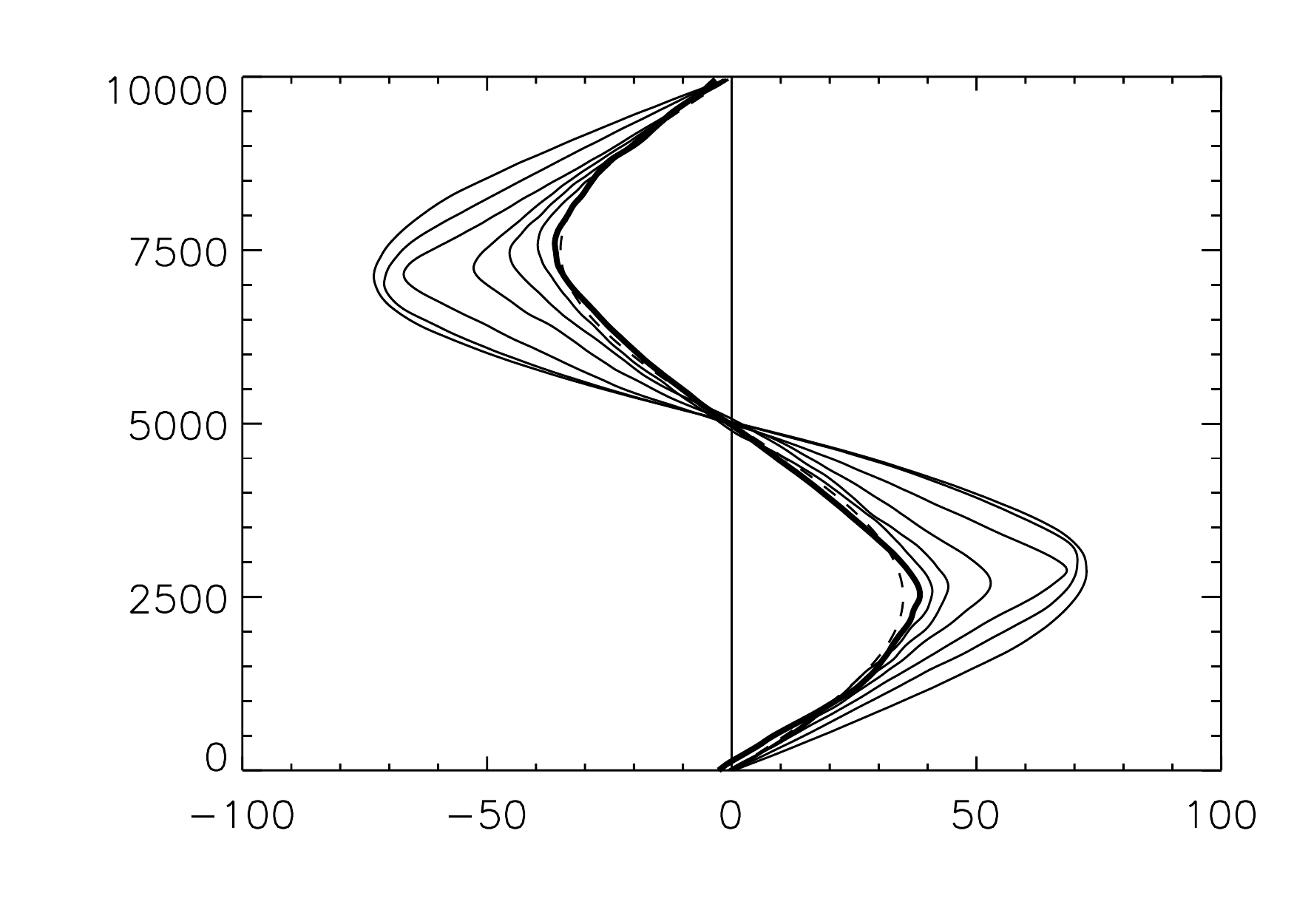}
\vspace{-1.2cm}
\caption{\footnotesize
Zonal-mean zonal velocity profiles $\uubar(y)$ in \ms\ ($y$ axis
in \km)
for the pure-DI case with
$\LD=1200$\km\ and $\qqstarmax=16$,
all shown
at time $t=120$ Earth years.
The inner, dashed curve is
$\uudeep(y)-U_0$.
The heavy solid curve is the upper profile $\uubar(y)$
for the anticyclones-only run,
$\fractionalbias = 1$ for all $\zeta$
(run \mbox{DI-12-16-1}).
The lighter solid curves show $\uubar(y)$,
in order of increasing peak $|\uubar|$,
respectively for
$\fractionalbiasmax=$ 1/4, 1/8, 1/16, 1/32, 1/64, and 0
(runs \,\mbox{DI-12-16-4}, \,\mbox{DI-12-16-8},...
\mbox{DI-12-16-$\infty$}).
}
\label{fig:pureDI-ubar-profiles}
\end{figure}

\begin{figure} [t]
\hspace{-0.50cm}
\includegraphics[width=22pc,angle=0]
{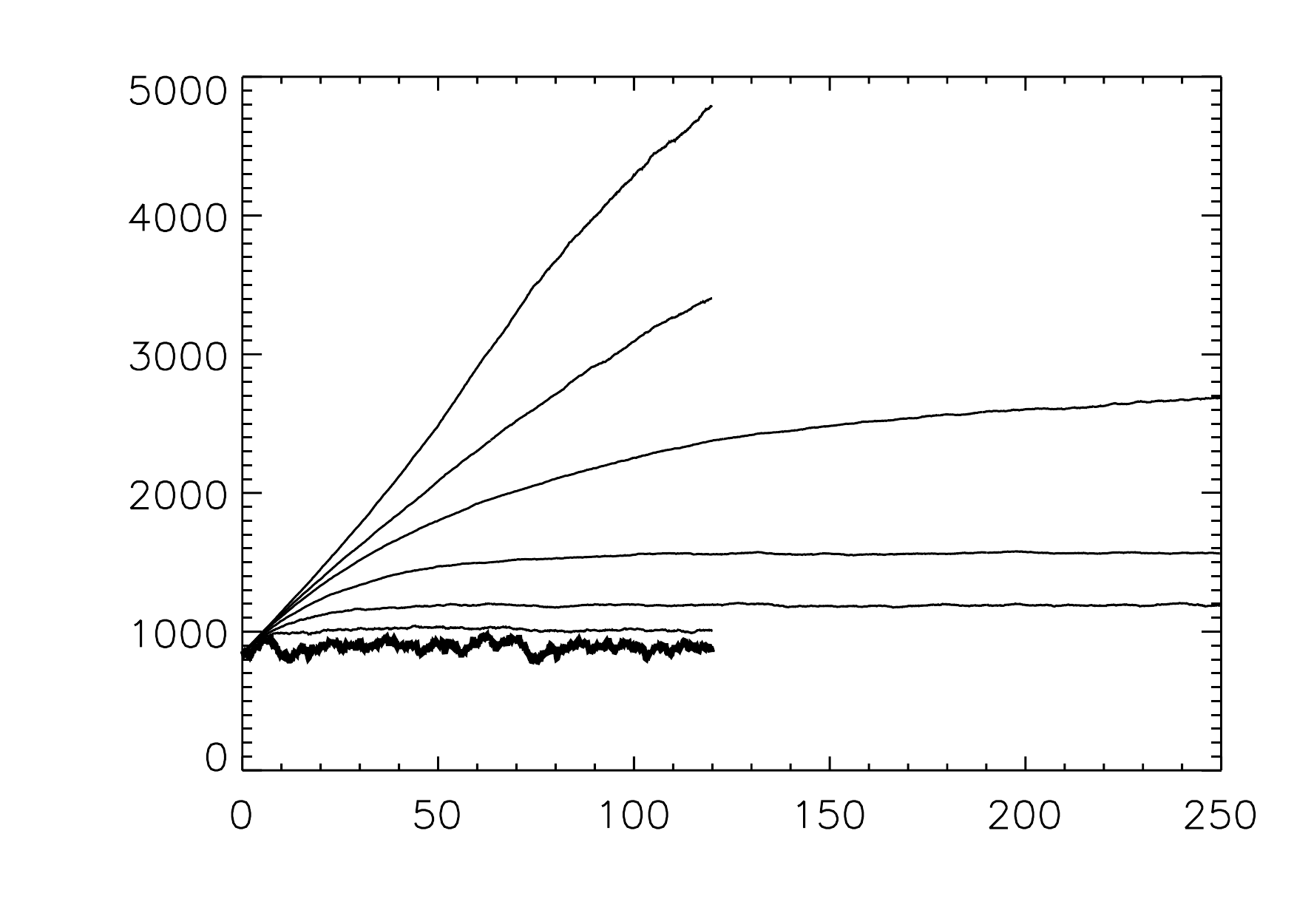}
\vspace{-1.2cm}
\caption{\footnotesize
Domain-averaged total energy per unit mass, in
\jkg\ or \mmss,
against time in Earth years for the
same set of pure-DI runs as in
Fig.~\ref{fig:pureDI-ubar-profiles},
with $\LD=1200$\km\ and $\qqstarmax=16$.
The lowest, heavy solid curve is for the anticyclones-only run,
$\fractionalbias = 1$ for all $\zeta$.
The lighter solid curves, reaching successively higher energies,
correspond respectively to
$\fractionalbiasmax=$ 1/4, 1/8, 1/16, 1/32, 1/64, and 0.
}
\label{fig:pureDI-energy}
\end{figure}

The behavior is roughly consistent with
the classic study of \cite{Kida1981}
on single vortices in shear.
In place of the
parabolic profile $\Delta\qq(\rr)$ defined by
(\ref{eq:injection-parabola}) above,
Kida's analysis assumes that
$\LD=\infty$ and uses a
top hat or ``vortex patch'' profile.
It should
be qualitatively relevant for small
vortex radius $\rr_0\ll\LD$.
Kida's condition for an anticyclone of strength
$\qqpatch$ to survive in cyclonic shear $S$
is $\qqstarpatch > 6.72$ where
$\qqstarpatch := |\sliver\qqpatch/S\sliver|$, as
can be shown straightforwardly from his
equations.\footnote{The
number $6.72$
can be verified from
the first line of Kida's Eq.~(3.4), by setting
$s = 1$
and plotting the right-hand side over an interval
$r\in(0, 1)$.
In Kida's notation $r=1$ corresponds to a
circular vortex and $r=0$ to a vortex
shredded by the shear
into an infinitely thin filament.
Taking $s=1$ picks out
the case of an initially circular vortex.
The vortex
is shredded
if for all $r$ the right-hand side of Kida's
(3.4) stays between $\pm1$, while if it dips below $-1$
the vortex survives.
For adverse pure shear such a dip occurs whenever, in Kida's notation,
$\omega/2e = -\omega/2\gamma > 6.7215$, where
$\omega=\qqpatch<0$
and $2\gamma=S>0$,
the shear.
For favorable pure shear
there is no sharp threshold,
but for instance shear with
$0< -\omega/2e = \omega/2\gamma < 0.5$
distorts
an initially circular vortex
beyond aspect ratio 25.
Its destruction is then a practical certainty
for finite $\LD$, or for almost any
background differing from Kida's
strictly steady, strictly constant pure shear of
infinite spatial extent.
  }
Model test runs with single-vortex injections
and vortex-pair injections
\citep[][Figs.~B10, B11 respectively]{mythesis}
behave
as expected from Kida's analysis.
For instance, a single parabolic anticyclone
of peak strength $|\sliver\qqpeaka|$
injected into cyclonic shear $S$
is destroyed when
$|\sliver\qqpeaka/S\sliver|\lesssim 7$.
It survives almost intact when
$|\sliver\qqpeaka/S\sliver|\gtrsim 16$,
then behaving almost like Kida's patch
with the same Kelvin circulation,
i.e., with the same value of
$2\pi\Antisliver\int_0^{\rr_0}\Antisliver\Delta\qq\Sliver\rr\sliver d\rr$
so that
$|\sliver\qqpatch| = \half|\qqpeaka|$.
For intermediate values
$8\lesssim|\sliver\qqpeaka/S\sliver|\lesssim 16$
 a small core survives while the
outskirts are eroded away.

The parameter $\psilim$ in (\ref{eq:def-psi-lim})
has to be chosen empirically.
We want the resulting $\zeta$ fields
to range over values within, or slightly exceeding,
the range  $-\half \leqslant\zeta \leqslant +\half$
that corresponds to
the sloping part of the ramp function $\ramp(\zeta)$.
A satisfactory choice is found to be
$\psilim=\Lambda\sliver\qqstarmax$
where $\Lambda=4.47\times10^6$\mms.  This is used in
all the runs mentioned here,
all the way from $\qqstarmax=0.5$ to $\qqstarmax=32$.
The precise value of $\Lambda$ is not critical.
Any neighboring value
will produce similar results.

\begin{table*}[t]
\centering
\begin{tabular}{|c|c|c|c|c|c|c|c|c|}
\hline
Run & $\beta_0$\footnotesize{/\smmone} & $\LD$\footnotesize{/\km}  & $\qqstarmax$  & $\text{Bias}$ & $\ttrun$\footnotesize{/Years} & Statistically& Figures\\
 & \footnotesize{Eq.(\ref{eq:def-betazero})ff.}  &
\footnotesize{Eq.(\ref{eq:pv-def})ff.} &
\footnotesize{Eq.(\ref{eq:qqstarmax-def})} &
\footnotesize{Eq.(\ref{eq:bias-defs})ff.} &  & steady? & \\
\hline
DI-12-16-1  & 0  & 1200  & 16  & 1  & 125  & Yes   & \ref{fig:pureDI-ubar-profiles}, \ref{fig:pureDI-energy}, \ref{fig:pureDI-qbar-profiles}, \ref{fig:pureDI-abar-profiles}\\
DI-12-16-4  & 0  & 1200  & 16  & 1/4  & 125  & Yes   & \ref{fig:pureDI-ubar-profiles}, \ref{fig:pureDI-energy}, \ref{fig:pureDI-qbar-profiles}, \ref{fig:pureDI-abar-profiles}\\
DI-12-16-8  & 0  & 1200  & 16  & 1/8  & 250  & Yes  & \ref{fig:pureDI-ubar-profiles}, \ref{fig:pureDI-energy}, \ref{fig:pureDI-qbar-profiles}, \ref{fig:pureDI-abar-profiles}\\
DI-12-16-16  & 0  & 1200  & 16  & 1/16  & 710  & Yes  & \ref{fig:pureDI-ubar-profiles}, \ref{fig:pureDI-energy}, \ref{fig:pureDI-maps-zeta-and-pv}, \ref{fig:pureDI-qbar-profiles}, \ref{fig:pureDI-abar-profiles}\\
DI-12-16-32  & 0  & 1200  & 16  & 1/32  & 710  & Yes  & \ref{fig:pureDI-ubar-profiles}, \ref{fig:pureDI-energy}, \ref{fig:pureDI-qbar-profiles}, \ref{fig:pureDI-abar-profiles}\\
DI-12-16-64  & 0  & 1200  & 16  & 1/64  & 125  & No  & \ref{fig:pureDI-ubar-profiles}, \ref{fig:pureDI-energy}, \ref{fig:pureDI-qbar-profiles}, \ref{fig:pureDI-abar-profiles}\\
DI-12-16-$\infty$  & 0  & 1200  & 16  & 0  & 125  & No  & \ref{fig:pureDI-ubar-profiles}, \ref{fig:pureDI-energy}, \ref{fig:pureDI-qbar-profiles}, \ref{fig:pureDI-abar-profiles}\\
ML-12-16-1  & $4.03\times 10^{-12}$ & 1200  & 16  & 1  & 125  & Yes  & \ref{fig:midlat-ubar-profiles}, \ref{fig:midlat-qbar-profiles}\\
ML-12-16-4  & $4.03\times 10^{-12}$ & 1200  & 16  & 1/4  & 125  & Yes  & \ref{fig:midlat-ubar-profiles}, \ref{fig:midlat-qbar-profiles}\\
ML-12-16-8  & $4.03\times 10^{-12}$ & 1200  & 16  & 1/8  & 250  & Yes  & \ref{fig:midlat-ubar-profiles}, \ref{fig:midlat-qbar-profiles}\\
ML-12-16-16  & $4.03\times 10^{-12}$ & 1200  & 16  & 1/16  & 290  & Yes  & \ref{fig:midlat-ubar-profiles}, \ref{fig:midlat-qbar-profiles}\\
ML-12-16-32  & $4.03\times 10^{-12}$ & 1200  & 16  & 1/32  & 240  & --  & \ref{fig:midlat-ubar-profiles}, \ref{fig:midlat-qbar-profiles}\\
ML-12-16-64  & $4.03\times 10^{-12}$ & 1200  & 16  & 1/64  & 125  & No  & \ref{fig:midlat-ubar-profiles}, \ref{fig:midlat-qbar-profiles}\\
ML-12-16-$\infty$  & $4.03\times 10^{-12}$ & 1200  & 16  & 0  & 125  & No  & \ref{fig:midlat-ubar-profiles}, \ref{fig:midlat-qbar-profiles}\\
DI-15-16-16  & 0  & 1500  & 16  & 1/16  & 295  & Yes  & \cite{mythesis}\\
ML-15-16-16  & $4.03\times 10^{-12}$ & 1500  & 16  & 1/16  & 60  & --  & \cite{mythesis}\\
DI-15-8-8  & 0   & 1500  & 8  & 1/8   & 60   & --   & \cite{mythesis}\\
DI-15-8-16  & 0  & 1500  & 8  & 1/16  & 320  & Yes  & \cite{mythesis}\\
ML-15-8-8   & $4.03\times 10^{-12}$ & 1500  & 8  & 1/8   & 60   & --  & \cite{mythesis}\\
ML-15-8-16  & $4.03\times 10^{-12}$ & 1500  & 8  & 1/16  & 315  & Yes & \cite{mythesis}\\
DI-12-8-16  &          0            & 1200  & 8  & 1/16  & 125  & --  & \cite{mythesis}\\
ML-12-8-16  & $4.03\times 10^{-12}$ & 1200  & 8  & 1/16  & 110  & --  & \cite{mythesis}\\
DI-12-32-16 &          0            & 1200  & 32 & 1/16  & 330  & Yes & \cite{mythesis}\\
DI-12-1-16  &          0            & 1200  & 1  & 1/16  & 350  & --  & \cite{mythesis}\\
DI-12-1-64   & 0  & 1200  &  1   & 1/64  & 300  & Yes  & \ref{fig:kelvin-ubar-profiles}, \ref{fig:kelvin-zetabar-profiles}, \ref{fig:kelvin-qbar-profiles}, \ref{fig:kelvin-abar-profiles}\\
DI-12-0.5-64 & 0  & 1200  & 0.5  & 1/64  & 300  & Yes  & \ref{fig:kelvin-ubar-profiles}, \ref{fig:kelvin-zetabar-profiles}, \ref{fig:kelvin-qbar-profiles}, \ref{fig:kelvin-abar-profiles}\\
\hline  
\end{tabular}
\vspace{0.2cm}
\caption{\footnotesize
Parameter values chosen.
In column~5, the bias values are
$\fractionalbias=1$ or $\fractionalbiasmax<1$;
see text surrounding Eq.(\ref{eq:fracbias}).
The
$\ttrun$ column
shows the
length of each run, rounded down to the nearest 5 
Earth
years.
Statistical steadiness is assessed at
$t=
\ttrun$.
The blank entries ``\;--\;''
signify runs not yet steady but judged likely to become so.
The parameter
$\psilim$ defined in
(\ref{eq:def-psi-lim}) is set equal to
$\Lambda\sliver\qqstarmax$
where
$\Lambda=4.47\times 10^6$\mms; $\Lambda$ is always held fixed.
Also,
$\ttmax$,
defined below Eq.~(\ref{eq:def-psideeptilde}),
is 24.06 hours in all runs except the last two,
\,\mbox{DI-12-1-64}\, and \,\mbox{DI-12-0.5-64},\,
for which $\ttmax=0.2406$ hours.
Other parameters held fixed are
$\Umax=35$\ms,
$L=1591.55$\km, $\rr_0=156.25$\km, $\Deltax = \Delta y$ =
39.0625\km, $\Deltat=50s$, $\Umax/L=2.199 \times 10^{-5}$\smone,
and the injected-vortex separation
$\s = 2\rr_0 + \Deltax= 351.56$\km,
Eq.~(\ref{eq:injection-separation}).
However, $\Deltax$, $\Delta y$ and $\Deltat$ are halved
in numerical resolution tests
\citep[e.g.,][\S3.5.2]{mythesis}.
}
\label{tab:parameter_values}
 \end{table*}

Table~\ref{tab:parameter_values} and its caption summarize
the parameter choices
for the most important runs.
In the first column, 
DI means ``pure
DI'' and ML ``midlatitude'', corresponding to the
$\beta_0$ values
shown in column~2.  See text
following (\ref{eq:def-betazero}).
The numbers within each label
\,\mbox{DI-12-16-1}\,, etc.,
are shorthand for $\LD$,
$\qqstarmax$, and bias respectively, as shown also in
columns~3--5.
Thus for instance \,\mbox{DI-12-16-1}\, labels a pure-DI run
in which
$\LD=1200$\km, $\qqstarmax=16$, and bias
$\fractionalbias = 1$, i.e.,
anticylones-only forcing.
When the final number exceeds 1, as in runs
\,\mbox{DI-12-16-4}\, to \,\mbox{DI-12-16-$\infty$},\,
it is the reciprocal of 
$\fractionalbiasmax$ in (\ref{eq:fracbias}).
In the column marked ``Statistically steady?'', the blank
entries ``\;--\;'' signify runs
not yet steady but likely to have become so, in our
judgement, had the run been continued for long enough.

\section{Main results}
\label{sec:mainresults}

\subsection{Pure-DI with \,$\LD=$1200\km, \,$\qqstarmax=16$,
 and varying bias}
\label{subsec:pureDI}

We focus at first on the pure-DI case with $\LD=1200$\km\
and $\qqstarmax=16$,
then comment briefly on
similarities and differences
for $\LD=1500$\km\
and for midlatitude cases.
Further details are given in \citet{mythesis}.
It is for $\LD=1200$\km, well below
the A2 stability
threshold, that we obtain the widest ranges of $\qqstarmax$
and $\fractionalbiasmax$ over which model flows are realistic
and statistically steady.
Broadly speaking,
the range of $\qqstarmax$ values that produce such flows
are found to be in or near
the semi-strong regime.

Figures~\ref{fig:pureDI-ubar-profiles}--\ref{fig:pureDI-abar-profiles}
show
results for the first seven runs in
Table~\ref{tab:parameter_values},
in which
bias is varied
in the pure-DI case
with $\LD=1200$\km\ and
$\qqstarmax=16$.

In Fig.~\ref{fig:pureDI-ubar-profiles}
the inner, dashed curve is the
deep-jet velocity profile
$\uudeep(y)-U_0$.
The solid curves are upper-jet profiles $\uubar(y)$
for
the
different biases, after 120\yr\ (Earth years)
of integration from the standard initialization.
The model belt
lies approximately in
the $y$-interval between 2500\km\ and 7500\km,
where the mean shears are cyclonic,
corresponding to
the central portion of
Fig.~\ref{fig:pureDI-ubar-profiles},
and of
Fig.~\ref{fig:cartoon} also.
The model zone is in the periphery and
its periodic extension.
The upper-jet profiles $\uubar(y)$
begin with the anticyclones-only run, \,\mbox{DI-12-16-1},\,
which has $\fractionalbias = 1$ for all $\zeta$.
This is the first solid curve, heavier than the rest and
only slightly different from the dashed curve.
The lighter
solid curves,
peaking at successively higher values of $|\uubar|$,
correspond to
runs \,\mbox{DI-12-16-4}\, to \,\mbox{DI-12-16-$\infty$},\,
i.e., to
$\fractionalbiasmax=$ 1/4, 1/8, 1/16, 1/32, 1/64, and 0
respectively.
We also ran $\fractionalbiasmax=1/2$;
the profile, not shown,
hardly differs from the dashed curve and the heavy,
$\fractionalbias=1$ profile.

Evidently the actual mean shears in the model belt
are either close to, or somewhat greater in
magnitude than, the nominal value $\Umax/L$
in (\ref{eq:qqstarmax-def}).
Many of the injections in these runs are semi-strong,
depending on injection locations.
A small minority can be strong.
The varying behavior of the injections
is further discussed below,
in connection with an illustrative movie.

As anticipated, reducing the bias reduces the
\qf al $\forcingbar$ effect,
allowing stronger upper jets.
In these pure-DI runs there is no dynamical difference between
prograde and retrograde jets, which on average are
sharpened and strengthened by
the same amounts.

The runs with
$\fractionalbiasmax$ ranging from $1/4$ to $1/16$,
and the run with $\fractionalbias=1$,
are all close to
statistical steadiness,
consistent with
the flattening-out of the corresponding curves
in Fig.~\ref{fig:pureDI-energy}.  These give
domain-averaged total energy
in \jkg\
against time $t$,
with bias
decreasing upward from curve to curve.
Total energies are dominated by
$\half |\uubar|^2 + \half\kd^2|\psibar|^2$,
the kinetic plus available potential energy
of the zonal-mean flow, contributing in roughly equal
proportions.
Domain-averaged
eddy energies, not shown, are relatively small
but also flatten out, for the runs in question.
The
run with $\fractionalbiasmax=1/32$
corresponds to the topmost of the three energy
curves that reach 250\yr.
It is evolving toward statistical
steadiness but does not come close to it until something like
500\yr\ of integration.
The run with $\fractionalbias = 1$,
included for comparison and
contrast with \citet{Li2006} and \citet{Showman2007},
is statistically steady apart from
a decadal-timescale
vacillation
(heavy curve at bottom of Fig.~\ref{fig:pureDI-energy}).
However, in that run the upper jets are hardly stronger than the
deep jets, as seen in Fig.~\ref{fig:pureDI-ubar-profiles},
and the $\zeta$ structure is correspondingly unrealistic.

\subsection{Details and movie for a realistic example}
\label{subsec:realistic-example}

\begin{figure} [t]
\hspace{-0.65cm}
\includegraphics[width=23pc,angle=0]
{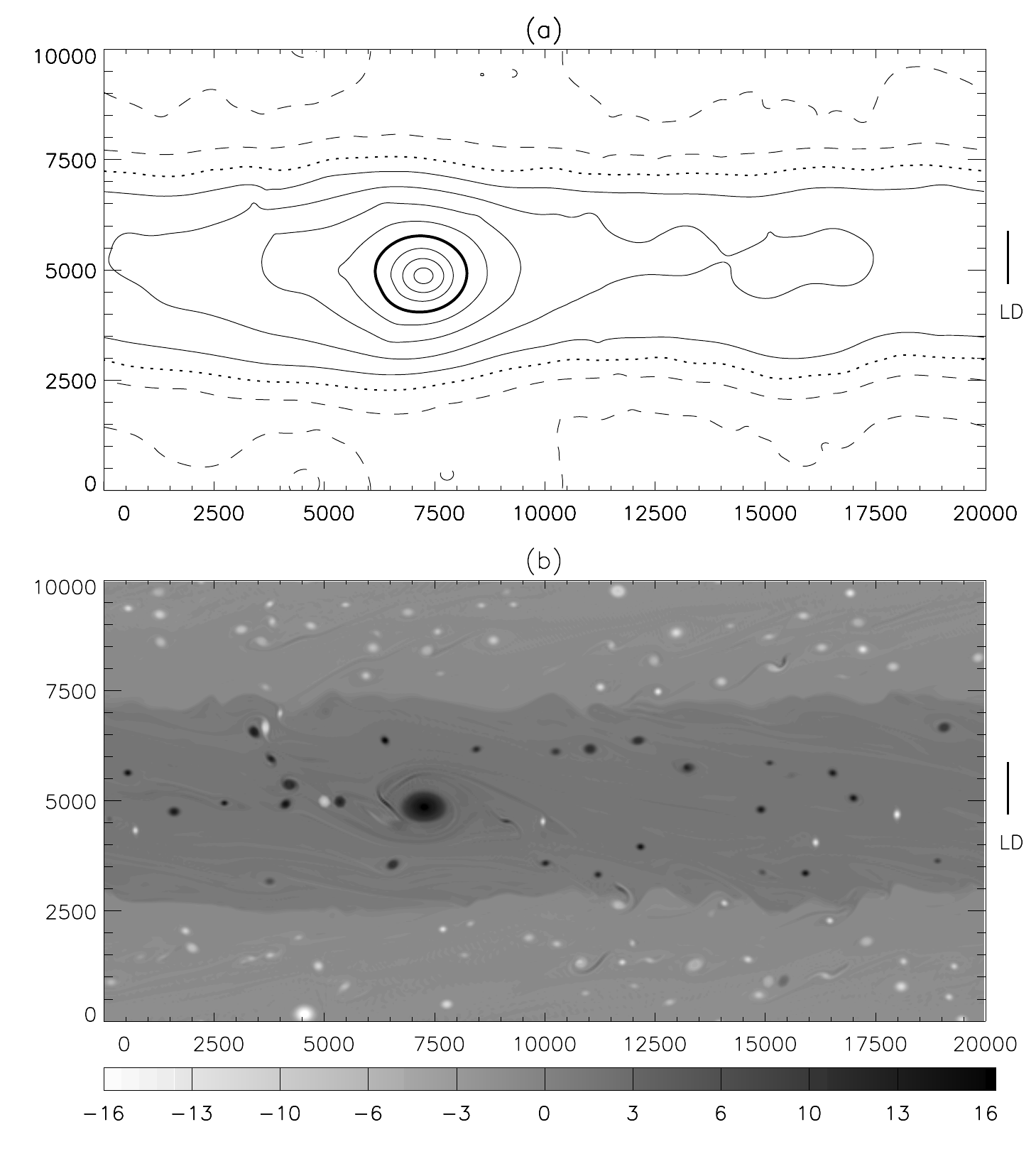}
\vspace{-0.8cm}
\caption{\footnotesize
Snapshots of the $\zeta$ and $\qq$ fields
at time $t=120$ Earth years for run \,\mbox{DI-12-16-16}
($x$ and $y$ axes in \km).
In the top panel (a), dashed contours show negative
$\zeta$ and solid contours positive
$\zeta$, with contour interval 0.1 in the dimensionless units
of Fig.~\ref{fig:injections-before-saturation}.
The heavy solid contour marks the value
$\zeta=+0.5$
at which the ramp function
$\rho(\zeta)$ saturates.
In the bottom panel (b),
which is the first frame of the supplemental movie,
the grayscale is in units of
$\Umax/L=2.199\times10^{-5}$\smone, like $\qqstarmax$.
The strongest vortices slightly exceed the grayscale range,
with the large cyclone in mid-belt, $y\simeq5000$\km,
peaking at $\qq=17.8\Umax/L$
and the largest anticyclone, to the far south-south-west
of it, near $x\simeq4500$\km, peaking
at $\qq=-18.4\Umax/L$.
There is one other out-of-range vortex, the small cyclone
north-north-west of the large cyclone,
near $x\simeq y\simeq6400$\km, which
peaks at $\qq=17.7\Umax/L$.
}
\label{fig:pureDI-maps-zeta-and-pv}
\end{figure}

We focus on run \,\mbox{DI-12-16-16}.
Fig.~\ref{fig:pureDI-maps-zeta-and-pv} shows snapshots
of $\zeta$ in contours and $\qq$ in grayscale
for that run,
at time $t=120$\yr.
A corresponding
$\qq$-field movie is
provided in the online supplemental material, in
grayscale and color versions.
The bars on the right show $\LD$.
Solid contours in Fig.~\ref{fig:pureDI-maps-zeta-and-pv}a\,
show positive $\zeta$,
a cold, elevated
interface
that
increases moist-convective activity.
The heavy solid contour
marks the value $\zeta=+\half$ at which the ramp function
$\rho(\zeta)$ saturates.  Dashed contours show negative
$\zeta$, a warm, depressed
interface
that
reduces moist-convective activity.
The structure of this $\zeta$ field is sufficiently zonal to
count as realistic, by our criterion that the model should
reflect the real planet's preference for
stronger convection in
belts
than in zones.

The $\qq$ snapshot
in Fig.~\ref{fig:pureDI-maps-zeta-and-pv}b
is dominated by small vortices, produced
by injections followed by
migration
 --- especially of small anticyclones from the belt
into the zone --- as well as by occasional vortex merging
and other interactions.
Cyclones are shown dark and anticyclones light.
The small
vortices move around chaotically,
under their neighbors' influence and that of the
background shear.
Yet vortex merging and upscale energy cascading
are inhibited to a surprising extent.
This lack of upscale cascading will be
discussed further in Section~\ref{sec:mechs-migration}.

Conspicuous in Fig.~\ref{fig:pureDI-maps-zeta-and-pv}b is
a single,
relatively large
cyclone
near $x\simeq$ 7500\km\ and $y\simeq$ 5000\km.
The strength
of this cyclone fluctuates but is statistically
steady.  The snapshot shows it
slightly larger and stronger
than average.
The strength is governed by reinforcement through
vortex merging
and internal migration on the one hand
(Section~\ref{sec:mechs-migration} below),
competing with attrition by
erosion and \qf al effects on the other.
The cyclone is strong enough
to produce a conspicuous footprint
in the $\zeta$ field, superposed on the quasi-zonal
structure (Fig.~\ref{fig:pureDI-maps-zeta-and-pv}a).
The footprint takes the form of a cold
patch or elevated
area extending
outward from the core of the
cyclone on the lengthscale $\LD$.

The velocity field of the large cyclone modifies the
background shear and strain quite substantially,
such that some of the nearby injection events
are strong rather than semi-strong.  An example can be seen in
the movie, starting west-south-west of the large cyclone
at $\ttrel=0.408$, where
time $\ttrel$ runs from 0 to 1
in units of movie duration, just under
an Earth month.
The injected anticyclone survives as it
travels around
the large cyclone, protected by the cyclone's
\emph{anticyclonic} angular shear,
then slowly migrates into the model zone to the north,
across the retrograde jet.
The accompanying cyclone, caught in the same
anticyclonic angular-shear environment,
suffers partial erosion almost immediately after injection.
It has a much shorter lifetime
and ends up completely shredded, at
$\ttrel\simeq0.58$, after one more partial erosion event.

Another clear example of migration from belt to zone,
this time southward,
occurs between $\ttrel\simeq0.65$ and $\ttrel\simeq0.9$.
A recently-injected
anticyclone partly merges with a pre-existing anticyclone,
near $x\simeq15000$\km\ and $y\simeq3000$\km,
and then migrates from belt to zone across the southern,
prograde jet.

The snapshot in
Fig.~\ref{fig:pureDI-maps-zeta-and-pv}b is
taken at the start of the movie, $\ttrel=0$.
At that instant, there has just been an injection
almost due west of the large cyclone, near
$x\simeq5000$\km.
That injection proves to be semi-strong.
Its anticyclone, seen on the left
in Fig.~\ref{fig:pureDI-maps-zeta-and-pv}b,
is shredded immediately.
However, its cyclone is also shredded shortly afterward,
again by the anticyclonic angular-shear environment.  During
the cyclone's short lifetime ($\ttrel\lesssim0.13$),
it migrates inward
through a small radial distance,
as can be checked by comparing its positions
south and then north of the large cyclone,
at $\ttrel = 0.040$ and $0.098$
respectively.  Such events are frequent and are clearly
part of what builds the large cyclone, whose
typical $\qq$-structure is sombrero-like,
a strong cyclonic
core surrounded by a weaker, fluctuating
cyclonic region,
easier to see in the color movie
than in the grayscale snapshot.
That structure is
alternately built up and eroded
by a chaotic sequence of vortex interactions and injections.

Also notable in the movie is
an injection making
a rare direct hit on the inner core of the large cyclone,
at $\ttrel=0.044$.
Thanks to the condition
(\ref{eq:saturation-constraint}) this injection behaves as a
semi-strong injection.
In this case the injected anticyclone is shredded into a tight spiral
and acts \qf ally.  By contrast,
the injected cyclone stays almost completely intact, and migrates
through a small radial distance
to the center.  The net effect is a slight
reduction in the overall
size and strength of the large cyclone,
from above average to below average.

\subsection{Varying bias again}
\label{subsec:other-biases}

Most of the statistically steady runs produce a
single, relatively large cyclone in a similar way.
Its average size increases as $\fractionalbiasmax$,
and hence
\qf, are reduced.
Runs \,\mbox{DI-12-16-64}\,
and  \,\mbox{DI-12-16-$\infty$},\,
with $\fractionalbiasmax=1/64$ and $0$, were terminated at 
$t=$ 125\yr\
because by then they had
developed single cyclones large enough
to produce an unrealistic, grossly nonzonal,
footprint-dominated
$\zeta$ structure.

The sharpened peaks of the jet profiles
for $\fractionalbiasmax=1/16$
in Fig.~\ref{fig:pureDI-ubar-profiles}
correspond to sharp steps
in the $\qq$ field, as seen in
Fig.~\ref{fig:pureDI-maps-zeta-and-pv}b
as sharp
transitions between light gray and darker gray.
These PV~steps,
embedded as they are
in relatively uni\-form surroundings,
resemble the cores of terrestrial strong jets
apart from their relatively limited meandering,
which is much more Jupiter-like.
The formation of such steps
from an initially smooth $\qq$ field
points to PV mixing
across belts and zones
as a contributing mechanism.
A role for PV mixing is consistent with the
chaotic appearance of the small-scale vortex interactions.

\begin{figure} [t]
\hspace{-0.50cm}
\includegraphics[width=22pc,angle=0]
{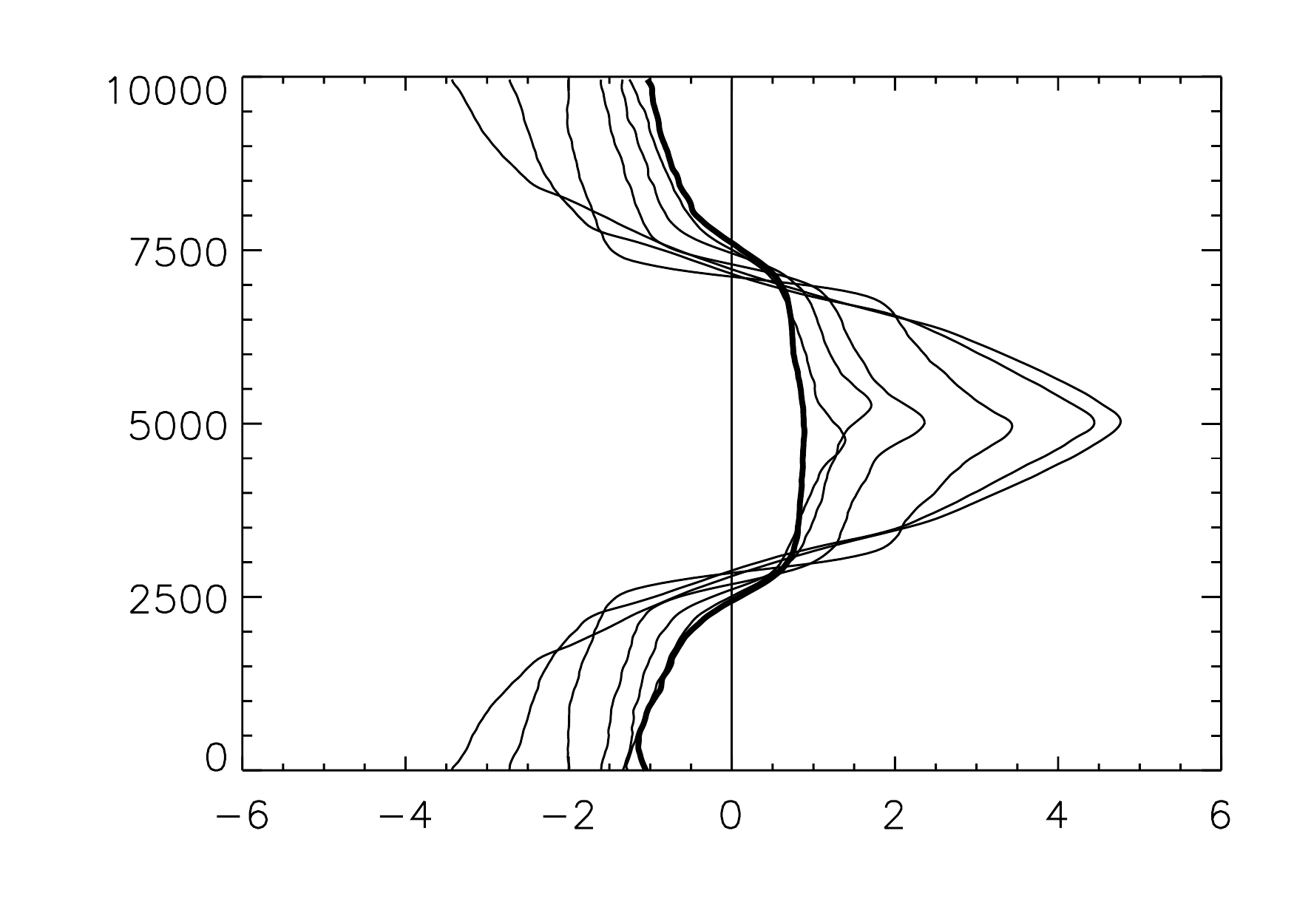}
\vspace{-1.2cm}
\caption{\footnotesize
Zonal-mean PV profiles $\qqbar(y)$ ($y$ axis in \km)
for the
same set of pure-DI runs as in 
Fig.~\ref{fig:pureDI-ubar-profiles},
with $\LD=1200$\km\ and $\qqstarmax=16$.  The PV is
in units of $\Umax/L=2.199\times10^{-5}$\smone\
and is time-averaged from $t=115$\yr\ to $t=125$\yr\
to reduce fluctuations.
The heavy curve is the $\qqbar(y)$ profile
for the anticyclones-only run,
$\fractionalbias = 1$ for all $\zeta$,
and the lighter curves with increasingly large
peak $|\qqbar|$ values
correspond respectively to
$\fractionalbiasmax=$ 1/4, 1/8, 1/16, 1/32, 1/64, and 0.
The initial profile, not shown, is sinusoidal
with amplitude 1 unit,
its central peak only just beyond the flat part of the
$\fractionalbias = 1$ heavy curve.
}
\label{fig:pureDI-qbar-profiles}
\end{figure}

\begin{figure} [t]
\hspace{-0.50cm}
\includegraphics[width=22pc,angle=0]
{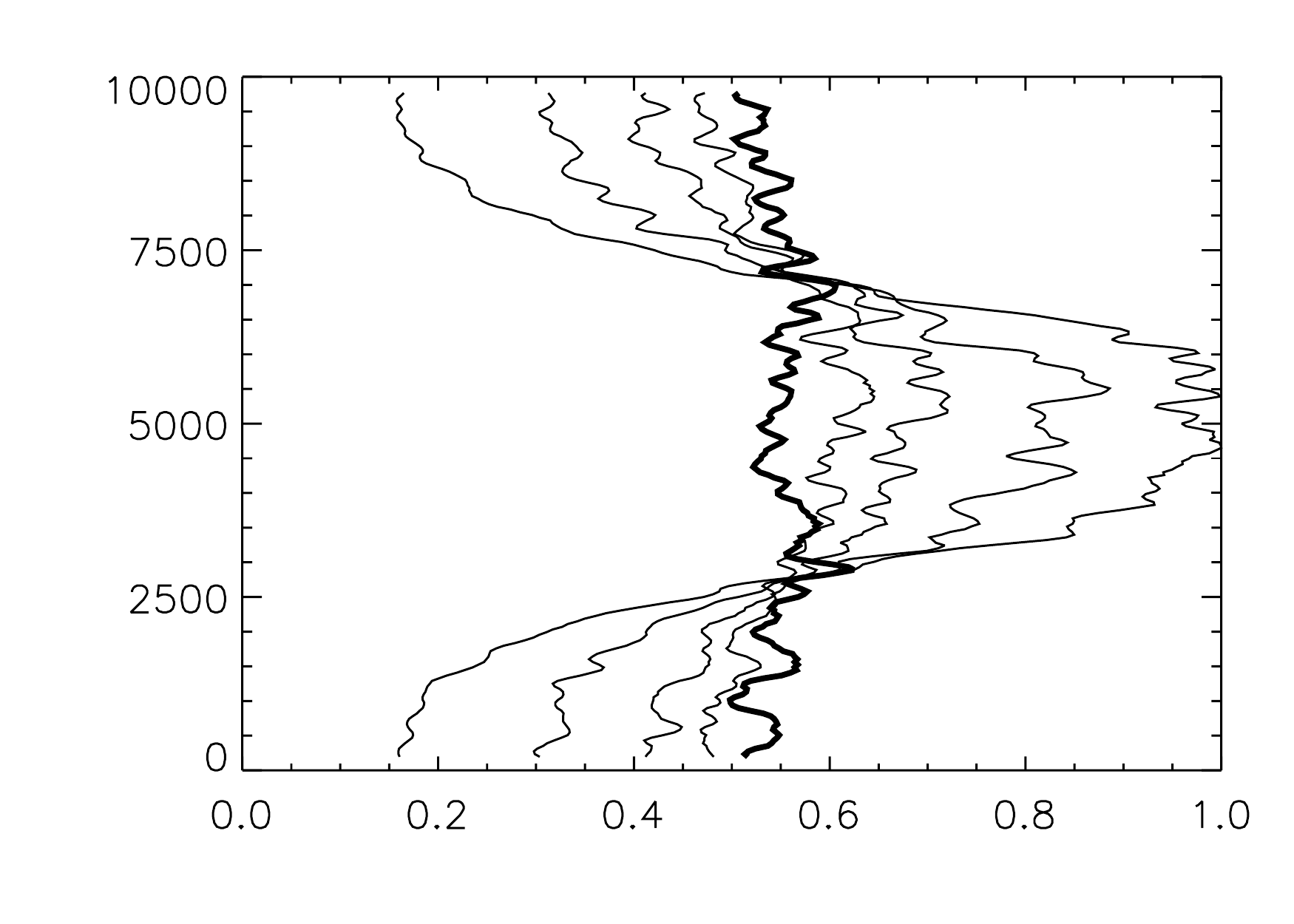}
\vspace{-1.2cm}
\caption{\footnotesize
Relative moist-convective activity
$\acforcing(y)/\acforcingmax$ ($y$ axis in \km)
for the
same set of pure-DI runs as in
Fig.~\ref{fig:pureDI-ubar-profiles},
except that $\fractionalbiasmax = 1/64$
        and $\fractionalbiasmax = 0$ are omitted.
Here
$\acforcing(y)$ is defined in Eq.~(\ref{eq:acforcing-def}),
and  $\acforcingmax$ is the largest value in the set shown.
The heavy curve is for
the anticyclones-only run $\fractionalbias = 1$.
The lighter curves are for
$\fractionalbiasmax=$ 1/4, 1/8, 1/16, and 1/32,
peaking successively further to the right.
The small wiggles arise from the statistical fluctuations
in the vortex-injection scheme,
showing up despite time-averaging from 60 to 120\yr.
}
\label{fig:pureDI-abar-profiles}
\end{figure}

\begin{figure} [t]
\hspace{-0.50cm}
\includegraphics[width=22pc,angle=0]
{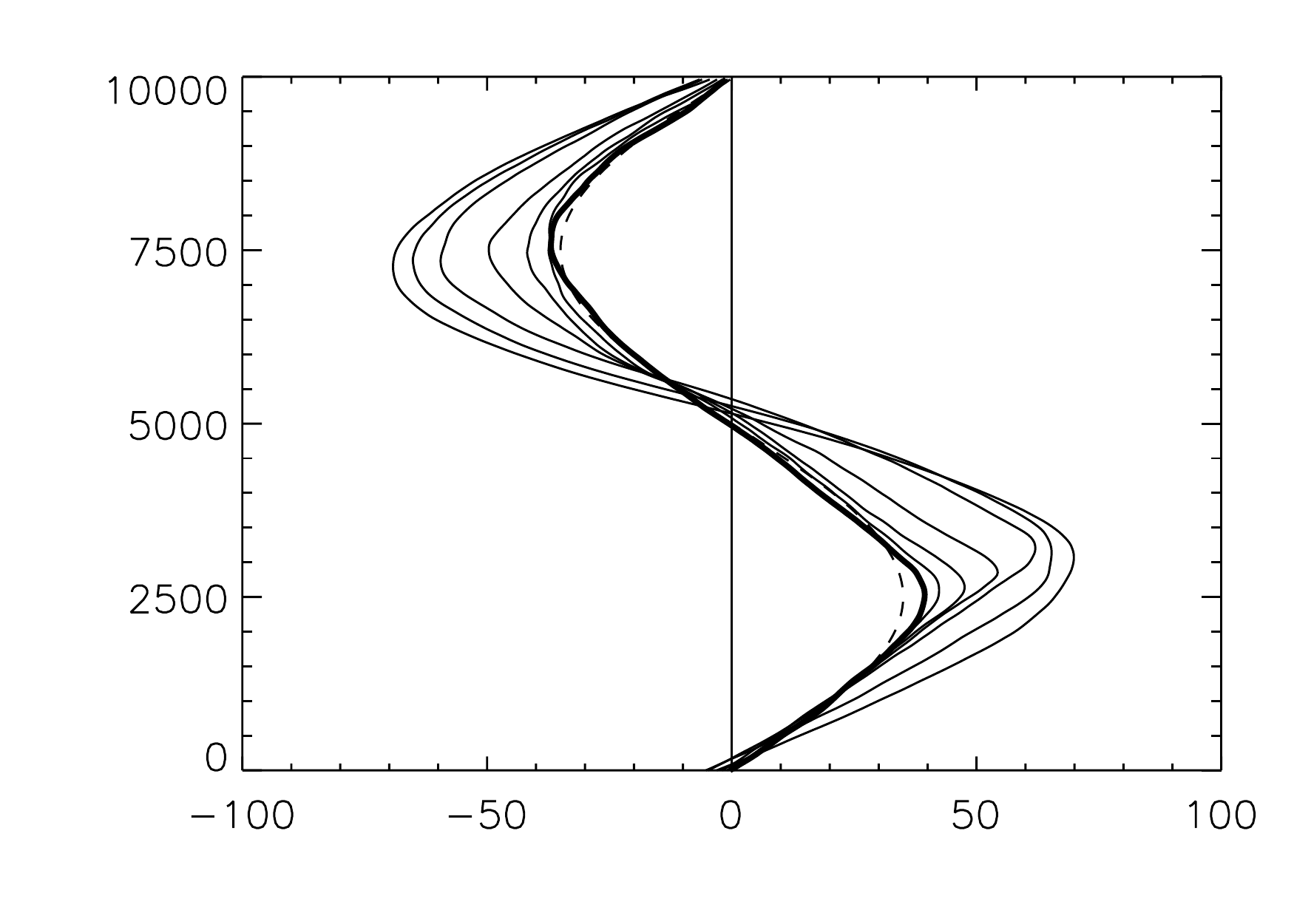}
\vspace{-1.2cm}
\caption{\footnotesize
Zonal-mean zonal velocity profiles $\uubar(y)$
in \ms\ ($y$ axis in \km)
for the midlatitude case with
$\LD=1200$\km\ and $\qqstarmax=16$,
at time $t=120$ Earth years.
Note that \emph{all} the retrograde jet profiles are rounded.
Biases vary as in
Fig.~\ref{fig:pureDI-ubar-profiles}.  These are the
runs labeled
\,\mbox{ML-12-16-1}, \,\mbox{ML-12-16-4},...
\mbox{ML-12-16-$\infty$}.
}
\label{fig:midlat-ubar-profiles}
\end{figure}

\begin{figure} [t]
\hspace{-0.50cm}
\includegraphics[width=22pc,angle=0]
{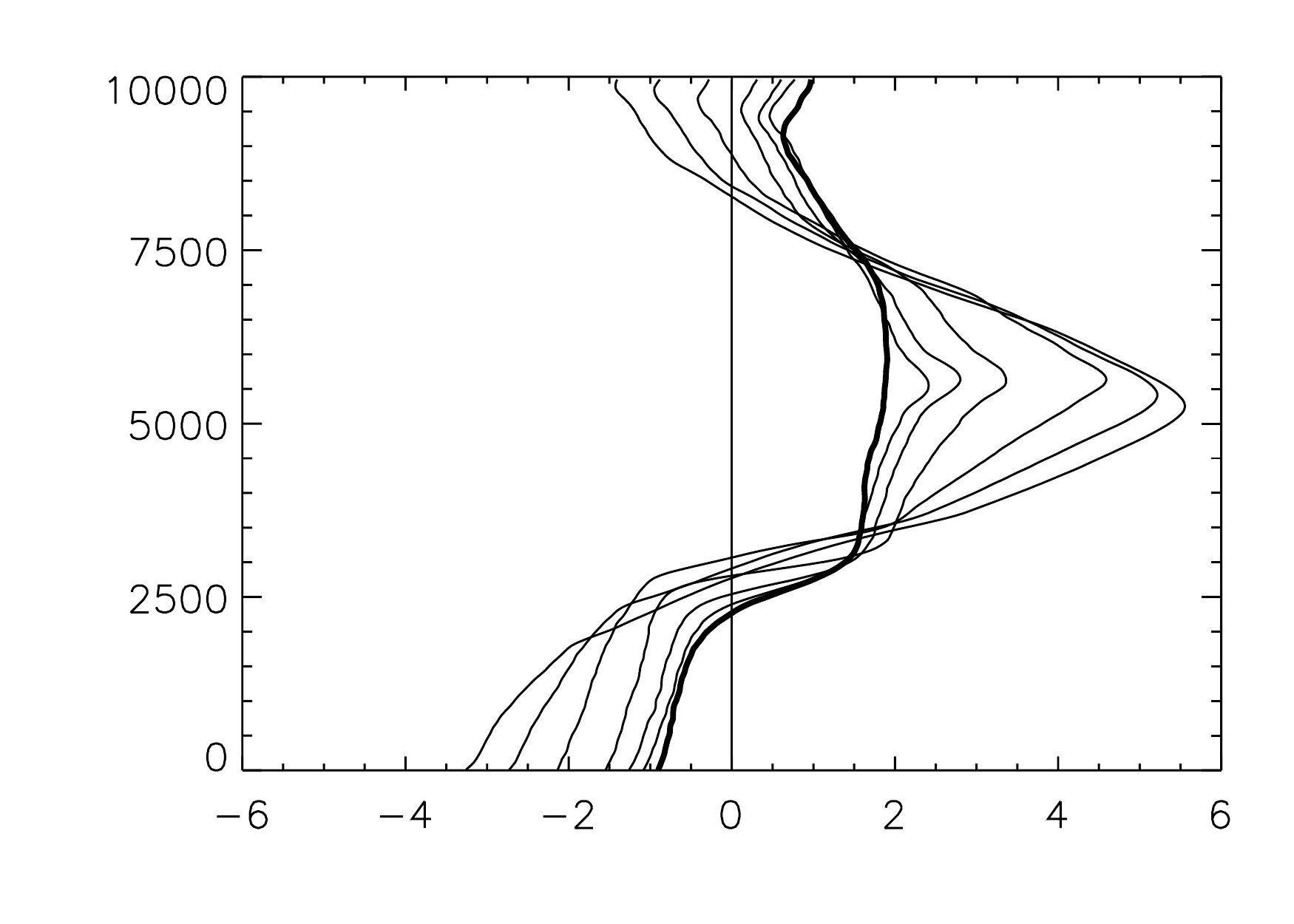}
\vspace{-1.2cm}
\caption{\footnotesize
Zonal-mean PV profiles $\qqbar(y)$ ($y$ axis in \km)
for the same set of midlatitude runs as in
Fig.~\ref{fig:midlat-ubar-profiles},
with
$\LD=1200$\km\ and $\qqstarmax=16$.  The PV is
in units of $\Umax/L=2.199\times10^{-5}$\smone\ and is
time-averaged from $t=115$\yr\ to $t=125$\yr\
to reduce fluctuations.
Biases
vary as in Fig.~\ref{fig:pureDI-qbar-profiles}.
The background PV gradient
$4.03\times 10^{-12}$\smmone\
makes each profile shear over,  with total displacement
$1.83$ units
($4.03\times 10^{-12}$\smmone\
$\times\;10^7$\m\
$\div\;2.199\times10^{-5}$\smone\
$=1.83$).
}
\label{fig:midlat-qbar-profiles}
\end{figure}

The PV~steps persist
into the two regimes with the smallest $\fractionalbiasmax$ values
1/64 and 0 and the largest cyclones.
However, the PV~steps
are no longer reflected in the corresponding $\uubar(y)$ profiles
in Fig~\ref{fig:pureDI-ubar-profiles},
the outermost two profiles.
Being Eulerian means, they are more rounded simply
because the large cyclones make the jets
meander more strongly.
A Lagrangian mean, not shown, would follow the meandering
and still reveal sharpened jet profiles --- indeed even
sharper than the sharpest in Fig.~\ref{fig:pureDI-ubar-profiles}.

Figure~\ref{fig:pureDI-qbar-profiles} shows
the Eulerian-mean $\qqbar$ profiles for
the same set of pure-DI runs,
at time $t\simeq120$\yr\ (see caption).
For $\fractionalbiasmax \geqslant 1/16$,
the profiles reflect the same
inhomo\-gen\-eous-PV-mixing structure,
though the large cyclone in
Fig.~\ref{fig:pureDI-maps-zeta-and-pv}b
creates a noticeable blip near $y\simeq$ 5000\km,
in the $\qqbar$ profile
for $\fractionalbiasmax = 1/16$.  Similar
blips, corresponding to
larger cyclones, become strong and then
dominant as $\fractionalbiasmax$ is reduced to 1/32, 1/64,
and 0; and the large cyclones are still growing in those
runs.  It is interesting to see what look like
hints of PV mixing
even for
$\fractionalbias = 1$, the heavy curve,
although the departure from the initial, sinusoidal
$\qqbar(y)$ profile is then very weak (see figure caption),
and unable to  produce
a realistic $\zeta$ field.

Figure~\ref{fig:pureDI-abar-profiles}
gives an alternative view of the model's
belt--zone structure
for $\fractionalbiasmax \geqslant 1/32$.
It shows in arbitrary units
a positive-definite measure $\acforcing$ of
average injection strength, defined by
\begin{equation}
\acforcing(y)
\;=\;
\overline{|\Delta\qq(\rr)|}^{xt}
\label{eq:acforcing-def}
\end{equation}
where the averaging is over many injections, both zonally and in
time.  In
Fig.~\ref{fig:pureDI-abar-profiles}
the time-averaging is from
60\yr\ to 120\yr.
Here
$\Delta\qq(\rr)$ represents the cores of injected anticyclones
only, as defined in
(\ref{eq:injection-parabola}),
i.e., omitting
their cyclonic
companions and complementary forcings.
Thus, $\acforcing=0$
would signal
a complete absence of injections.
Increasing $\acforcing$-values signal increasingly strong
injections, on average, regardless of bias.
The wiggles in the curves arise from the statistical fluctuations
in the vortex-injection scheme.

The belt-to-zone
variation in injection strength
seen in Fig.~\ref{fig:pureDI-abar-profiles}
is consistent with the realistic $\zeta$ structures
found for moderately small values of $\fractionalbiasmax$,
with the strongest injections concentrated in mid-belt.
The unrealistic $\zeta$ structure for $\fractionalbias = 1$
produces, as expected,
relatively little belt-to-zone variation in injection strength
(heavy curve).
The belt-to-zone variation
increases as $\fractionalbiasmax$
decreases toward 1/32, but then decreases again (not shown)
because of the dominance of the large cyclone,
within whose footprint
the condition (\ref{eq:saturation-constraint})
weakens the injections.

\subsection{Other cases and parameter values}
\label{subsec:other-cases}

Results for $\qqstarmax=8$
\citep[not shown; see][]{mythesis}
show similar behavior, though the tendency to form large
cyclones is weaker,
and there are cases in which the largest cyclones
come in pairs.
Most injections are then weak or
semi-strong.  For $\qqstarmax=32$, by contrast, many injections are
strong, resulting in relatively violent vortex activity.
A long run \,\mbox{DI-12-32-16}\, has
been carried out for $\qqstarmax=32$ and $\fractionalbiasmax=1/16$.
It shows unrealistic, strongly
nonzonal $\zeta$ structure,
briefly described in
Section~\ref{sec:mechs-migration}\ref{subsec:large-anticyc}.
For $\qqstarmax=1$ or less, practically all injections are weak.
The response is then governed mainly by the
Kelvin and $\forcingbar$ mechanisms.  See
Section~\ref{sec:kelvin}.

For the ``midlatitude'' case, with
$\beta_0 = \beta - \kd^2U_0=4.03\times 10^{-12}$\smmone,
the value of $\beta$ itself
at latitude 35\degree,
the results \citep{mythesis} are broadly similar except that
jet-sharpening is more effective for the
prograde than for the retrograde jets.
The $\uubar$ and $\qqbar$ profiles for
$\LD=$ 1200\km\ and $\qqstarmax=16$ at time
$t\simeq120$\yr\
are shown in
Figs.~\ref{fig:midlat-ubar-profiles}
  and~\ref{fig:midlat-qbar-profiles}
(see captions).
Again, only the runs with
$\fractionalbias = 1$ or
$\fractionalbiasmax\geqslant1/16$
are close to statistical steadiness at
$t\simeq120$\yr.

Notice from Fig.~\ref{fig:midlat-qbar-profiles}
that the $\qqbar$ profiles are still strongly nonmonotonic.
The $\zeta$ and $\qq$ fields
for $\fractionalbiasmax=1/16$
are similar
to those in Fig.~\ref{fig:pureDI-maps-zeta-and-pv},
except that the
PV~step
near $y=$ 7500\km\
is distinctly weaker
and the large cyclone distinctly stronger.
Concomitantly,
the $\zeta$ field is somewhat less zonal,
with a stronger and larger cyclonic footprint.

The pattern of results for $\LD=$ 1500\km\ is
again broadly similar,
except that realistic quasi-zonal
structure is more easily disrupted.
Vortex interactions reach across somewhat greater distances,
and the whole system is somewhat closer to
the A2-marginal stability threshold.
This makes the upper
jets more liable to large-scale meandering.
The most
realistic $\zeta$ fields are obtained for a narrower range of
$\qqstarmax$ values, closer to 8 than 16.
For further detail see \citet{mythesis}.

\subsection{Regarding statistical steadiness}
\label{subsec:statistical-steadiness}

As an extreme test of statistical steadiness,
we ran our main case \,\mbox{DI-12-16-16}\, out to
$t=710$\yr\
and compared details at later times with the
120-\yr\ results shown above.
All the mean profiles remain nearly indistinguishable
from those in Figs.~\ref{fig:pureDI-ubar-profiles},
\ref{fig:pureDI-qbar-profiles}, and
\ref{fig:pureDI-abar-profiles}.
The $\qq$ fields are
qualitatively indistinguishable from
Fig.~\ref{fig:pureDI-maps-zeta-and-pv}b.
In particular, the
largest cyclone and anticyclone have similar sizes.
By way of illustration a snapshot of the $\qq$ field
at 600\yr\ is shown in \citet{mythesis},
Fig.~4.10.

As a further check, we produced a time series of domain-maximum
cyclone strength over the whole time interval from $t = 0$ to
710\yr.
The time series showed fairly strong fluctuations,
mostly in the range 15--20 in units of $\Umax/L$.
Many of these fluctuations are fleeting and are, we think, due to
transient Gibbs fringes
produced by the high-wave\-num\-ber filter,
during interactions involving the strongest small cyclones.
The main point, however, is that
even this sensitive
time series looks
statistically steady from
$t\simeq100$\yr\ onward,
all the way out to
$t=710$\yr.

\section{Mechanisms in play}
\label{sec:mechs-migration}

\subsection{To merge or not to merge}
\label{subsec:merge}

Repeated
vortex merging, giving rise to an
upscale energy cascade
as vortices get larger --- a stepwise energy transfer in wavenumber
space ---
is often taken for granted
as an essential mechanism in all two-dimensionally turbulent flows.
It therefore came as a surprise to us to discover that the
cascade mechanism plays
no significant role at all,
in our most realistic cases.

In these cases
the stochastic forcing is
strong enough to
produce nonlinear behavior in the form of
chaotic vortex interactions.   The
vortices follow trajectories resembling random walks,
encountering each other in
a quasi-random way as is evident from
$\qq$-field movies such as that discussed in
Section~\ref{sec:mainresults}\ref{subsec:realistic-example}.
Vortex-merging events do occur,
as noted in that discussion,
but require extremely close encounters and are
much rarer than vortex erosion events.
Except for one or two large cyclones, such as the single large
cyclone in Fig.~\ref{fig:pureDI-maps-zeta-and-pv}b, all the
other vortices remain small, in all the realistic cases
we have seen --- in order-of-magnitude
terms not much larger than
the vortices
originally injected.
The largest anticyclone in Fig.~\ref{fig:pureDI-maps-zeta-and-pv}b,
at the bottom of the figure near
$x\simeq4500$\km,
shows that
whatever merging events took place,
during the preceding 120\yr,
they could not have been enough to produce
much systematic growth in vortex size despite the favourable
background shear.\footnote{A
reviewer asks for more evidence
against vortex merging and upscale energy cascading.
Our best answer, beyond saying that the
$\qq$-field
movie described in
Section~\ref{sec:mainresults}\ref{subsec:realistic-example}
shows typical behavior
--- as also seen in the output from other runs ---
is to call attention to the extremely long run described in
Section~\ref{sec:mainresults}\ref{subsec:statistical-steadiness}.
Even after 710\yr\ of integration, vortex merging has made no
net headway against vortex erosion and shredding; and the
distribution of vortex sizes remains qualitatively
indistinguishable from that in
Fig.~\ref{fig:pureDI-maps-zeta-and-pv}b,
including the size of the largest anticyclone.
That is, all the
vortices remain small in comparison with $\LD$ apart from
the single large cyclone, which persists as a coherent entity
whose size is statistically steady.  The other point to make is
that, as will be noted in
Sections~\ref{sec:mechs-migration}\ref{subsec:eddyfluxes}--\ref{sec:mechs-migration}\ref{subsec:noncascade},
the upscale energy transfer responsible for
strengthening the upper jets depends mainly on vortex migration,
with little dependence on vortex merging, and is a direct and not a
stepwise transfer from small scales to large.
}

A contributing factor
is the sparseness of our vortex injection scheme,
idealizing the intermittency of real Jovian moist convection
as in \citet{Li2006}.
This contrasts with the extremely dense forcing --- dense both
spatially and temporally --- used
in orthodox beta-turbulence studies.
Such studies use a spatially dense forcing
of a special kind, in order
to achieve spectral narrowness
\citep[e.g.,][Eq.~(20) and Fig.~1f]{Srinivasan2014}.

Sparse forcing need not, by itself, lead to sparse vortex fields,
in a model with numerical dissipation small enough to
allow long vortex lifetimes.
For isolated strong vortices
having our standard size
$\rr_0=4\Deltax$, injected into favorable shear,
and with peak
strength 16 times the shear,
we find that lifetimes under numerical dissipation alone are
typically of the order of years, albeit
variable because they depend on the Robert filter and on
bulk advection speeds across the grid
\citep[][\S2.1.1]{mythesis}.

For comparison, average injection rates
are of the order of
2~vortex pairs per day
in all the cases
described above; and so
the modest number of small vortices seen in snapshots like
Fig.~\ref{fig:pureDI-maps-zeta-and-pv}b
can be explained only if
vortex lifetimes are more like
months than years.
Vortex lifetimes are therefore
limited not by numerical dissipation but by
the chaotic vortex interactions themselves, as
illustrated by the erosion and shredding
events seen in the $\qq$-field movies.

The longest-lived small vortices are the anticyclones in the
zone.
Of these, the weakest come from local injections,
corresponding to low values of the ramp function
$\ramp(\zeta)$, and the strongest from migration events
like the two events
described
in Section~\ref{sec:mainresults}\ref{subsec:realistic-example}.
Such migration, of small but relatively strong
anticyclones from belt to zone, can be attributed to a
combination of chaotic,
quasi-random walking
away from strong-injection sites on the one hand,
and the so-called ``beta-drift'' or ``beta-gyre''
mechanism on the other.

\subsection{The beta-drift or beta-gyre mechanism}
\label{subsec:betadrift}

As is well known, and as we have verified by
experimentation with our model, a single vortex injected into a
background PV gradient will immediately advect the background gradient
to produce a pair of opposite-signed PV anomalies on either side,
traditionally
called ``beta-gyres'',
whose induced velocity field advects the original vortex toward
background values closer to its own PV values.  This migration
mechanism weakens as the anomalies wrap up into a spiral pattern
around the original vortex.  Nevertheless, the mechanism appears to
have a role in helping an anticyclone to cross
a jet, from belt to zone either northward or southward.
Such an anticyclone
typically carries with it a wrapped-up cyclonic fringe,
which is subsequently eroded away.

The inward migration of injected \emph{cyclones}
within a larger cyclone
plays a role in the buildup and persistence of
the statistically-steady
large cyclones we observe.  For instance
the example in the movie between $\ttrel\simeq0.04$ and $0.10$
(Section~\ref{sec:mainresults}\ref{subsec:realistic-example})
does, on close inspection,
show a local beta-drift mechanism
in operation, the neighboring PV contours being
weakly twisted in the sense required, as is especially clear
around $\ttrel\simeq0.10$.
The corresponding mechanism for large anticyclones seems
too weak to compete with erosion, in the regimes explored so far
that have realistic $\zeta$ structure.

\subsection{The eddy fluxes
            $\overline{\vv'\qq'}$ and  $\overline{\uu'\vv'}$}
\label{subsec:eddyfluxes}

For brevity, no
profiles of $\overline{\vv'\qq'}$ and
$\partial(\overline{\vv'\qq'})/\partial y$ are
shown here.
Their qualitative characteristics
are, however, simple, and easy to see
for realistic, statistically steady states like that of
Fig.~\ref{fig:pureDI-maps-zeta-and-pv}.
For then, broadly speaking,
$\zeta$ and $\fractionalbias$ are
maximal in mid-belt and minimal in mid-zone,
having regard to (\ref{eq:fracbias})
and to quasi-zonal
$\zeta$ fields like that in
Fig.~\ref{fig:pureDI-maps-zeta-and-pv}a.
In a statistically steady state the right-hand side of
(\ref{eq:mean-pv-equation}) must vanish, after
sufficient time-averaging.
Thus averaged,
$\partial(\overline{\vv'\qq'})/\partial y$ must therefore
have the same $y$-profile as $\forcingbar$.
Apart from a sign reversal
and an additive constant,
such $\forcingbar$ profiles are shaped like
smoothed versions
of the profiles of $\acforcing$
in Fig.~\ref{fig:pureDI-abar-profiles}, in realistic cases,
although $-\forcingbar$ tends to be
more strongly peaked in mid-belt because of the dependence
(\ref{eq:fracbias}) of $\fractionalbias$ upon $\zeta$.
The peak in $-\forcingbar$ comes from injections that are both
stronger, as indicated by the mid-belt peak in
$\acforcing(y)$, and more strongly biased because
$\fractionalbias$ itself peaks in mid-belt.
The additive constant is
required in order to make
$\int_0^{2\pi L}\antisliver\forcingbar\,dy=0$
for consistency with (\ref{eq:domain-ints-zero}b).

Periodicity and the Taylor identity (\ref{eq:taylor-identity})
require, moreover, that
$\int_0^{2\pi L}\antisliver\overline{\vv'\qq'}\,dy=0$.
The $\overline{\vv'\qq'}$ profile therefore has to be
qualitatively like an
additive constant plus
the periodic part of the indefinite integral of
$-\sliver\acforcing(y)$.
Again after sufficient
time-averaging
this is a smooth, quasi-sinusoidal curve going
positive to the south and negative to the north of mid-belt,
which is consistent with the
migration of
strong anticyclones
from belt to zone already noted.
Both $\overline{\vv'\qq'}$ and $\overline{\uu'\vv'}$ are
upgradient
at nearly all latitudes,
in pure-DI cases.
Explicit diagnostics of the model output confirm this
qualitative picture \citep[][Figs.\ 4.12, 4.13]{mythesis}.
There is no PV-mixing signature
in the statistically steady state
essentially
because, as illustrated in
Fig.~\ref{fig:pureDI-maps-zeta-and-pv}b,
the mixing has already taken place.

\subsection{Non-cascade energy transfer and
                          \qf al timescales}
\label{subsec:noncascade}

The upgradient migration of small anticyclones from belts
to zones
is crucial to the strengthening of the upper jets
relative to the deep jets.  It is therefore crucial to
obtaining realistic 
belt--zone contrasts in small-scale convective activity.

The migration produces a clustering of anticyclones, on the largest
possible meridional scale, into regions whose background is
already anticyclonic.
Energetically, this of course implies an
energy transfer from
vortex-injection scales
directly to the largest meridional
scales, as distinct from a Kolmogorovian turbulent energy
cascade via intermediate
scales.  The direct transfer is balanced by the 
\qf al $\forcingbar$ effect,
taking the large-scale energy back out of the system.

\Qf al
timescales in our model are of the order of
decades,\footnote{For
instance run \,\mbox{DI-12-16-16}\, has a \qf al
timescale $\sim 15$\yr, as can be estimated from the steady-state
$\forcingbar$ profile for that run in Fig.~4.12d of
\citet{mythesis}, peaking at just over $4\times10^{-8}$\mss,
and the
difference between
the $\fractionalbiasmax=1/16$ curve and
the dashed curve
in Fig.~\ref{fig:pureDI-ubar-profiles} above,
peaking at just under 20\ms.
}
considerably shorter than radiative timescales near the bottom
interface, which even without clouds can be estimated as
more like a century
(e.g.,
\citeauthor{Gierasch1969}
\citeyear{Gierasch1969};
\citeauthor{Pierrehumbert2010}
\citeyear{Pierrehumbert2010}, Eq.~(4.24);
Roland Young, personal communication).

\subsection{The large vortices
 in run DI-12-32-16}
\label{subsec:large-anticyc}

The run \,\mbox{DI-12-32-16}\, 
\citep[for details again see][]{mythesis}
develops not only a large cyclone but also a large anticyclone,
perhaps reminiscent of the real planet's Ovals,
though more symmetrically located
within the model zone.
There are two
caveats about this run.
First and most important,
the accompanying $\zeta$ structure
is footprint-dom\-in\-ated, and not
quasi-zonal
as in Fig.~\ref{fig:pureDI-maps-zeta-and-pv}a.
So we count it
as unrealistic.
The model's
large anticyclone depends
less on belt-to-zone migration than on
strong injections directly into the zone.

Second, the two large vortices and their periodic images
form a vortex street,
more precisely a doubly-periodic
vortex lattice,
constrained by the 2:1 geometry of the
model domain.  Without extending our study to a much larger domain
we cannot, therefore,
claim to be capturing possible vortex-street properties in any
natural way.

\subsection{Runs with no deep jets}
\label{subsec:nodeepjets}

It might be asked what happens to flow regimes like that of
Fig.~\ref{fig:pureDI-maps-zeta-and-pv},
run DI-12-16-16, when the deep jets are removed
keeping everything else the same. Part of the answer is clear from
inspection of the A2 stability criterion (\ref{eq:a2-stability}).
For our standard
pure-DI initial conditions, with sinusoidal $\uubar(y)$, removing
the deep jets puts the flow well above the shear-stability
threshold.

Indeed (\ref{eq:a2-stability})
then collapses to  $\kkmin^2 > L^{-2}$ as the condition for stability.
Thus the stability threshold becomes independent of $\LD$, and the
initial jets are always unstable in the 2:1 domain or in any
other doubly-periodic domain whose zonal extent exceeds the jet
spacing.  The counterpropagating Rossby waves can no longer be
decoupled by reducing $\LD$.
Reducing $\LD$
produces a precisely
compensating increase in  $\partial\qqbar/\partial y$  through the term
$\kd^2\uubar$ in (\ref{eq:mean-pv-grad}).
For our standard pure-DI initial conditions,
therefore, the stabilization hence straightness of the upper
jets depends on the presence of the deep jets as well as on
having an $\LD$ value that is not too large.

Runs DI-12-16-16 and DI-12-16-1 were repeated with the deep jets
removed, that is, with $\Umax = 0$ in (\ref{eq:deep-jets}).
As expected, the initial flows were
then violently unstable to the longest available waves,
those with wavenumber $\kkmin$.  The initial zonal jets were
rapidly destroyed, leaving behind a quasi-random
distribution of wandering
vortices of various sizes.
These included some very large vortices,
one cyclone and one anticyclone per domain cell, in the first of
the two runs.  In that run
their sizes were comparable to the sizes of the large vortices
in the run DI-12-32-16 described in the last
subsection.  However, their behavior was quite different.
They wandered quasi-randomly
rather than being locked
into a stable doubly-periodic lattice.  It seems that 
in run DI-12-32-16 the influence of the deep jets was essential
to stabilizing the lattice of large vortices.

The second run, DI-12-16-1 with its deep jets removed, began
similarly with a violent long-wave instability
but then settled into a state whose
largest vortices, all anticyclones in this case, were somewhat
smaller in size and greater in number.
The behavior was
quite unlike that found by \citet{Li2006}.
We may note that Li et al.\ studied regimes 
with $\fractionalbias=1$ and no deep jets but also
with strong monotonic background PV
gradients.  Indeed their case having the most realistic jet speeds
(their Fig.~9) had a $\beta_0$ value significantly enhanced by
introducing a strong, 80\ms, \emph{retrograde} deep flow $U_0 < 0$;
recall (\ref{eq:def-betazero}).

To better understand Li et al.'s results, a set of runs
was carried out with $\fractionalbias=1$,
monotonic background PV, no deep flow
($\Umax = U_0 = 0$),
and other choices aimed at getting
as close as possible to
the conditions they studied; for more detail see
\citet[][\S3.5.1 and Figs.\ B.1--B.9]{mythesis}.
Besides the large \qf\ for $\fractionalbias=1$,
two other mechanisms were found to be important, first,
inhomogeneous
PV mixing, converting the background PV gradient
into a fairly sharp staircase giving sharp prograde jets
and, second, predominantly
equatorward migration of the injected anticyclones through the
(still-monotonic) background PV gradient.  In our doubly-periodic
domain, this had a new effect not reported  
in Li et al.'s channel-domain results.
The migration,
predominantly equatorward out of
the belts,
produced a persistent
phase shift in $\qqbar(y,\,t)$ that gave rise
to persistent
poleward propagation of the jet system.
That propagation is perhaps reminiscent of the downward
propagation of the terrestrial quasi-biennial oscillation,
though we emphasize that its mechanism is quite different.
In any case, no
poleward
propagation of the jet system is
seen on the real planet.

\section{The Kelvin mechanism}
\label{sec:kelvin}

The Kelvin/CE2/SSST passive-shearing mechanism has gained
increased attention recently
\citep[e.g.,][\andrefs]{Srinivasan2014}.
It is one of three
very different
mechanisms for creating and sharpening jets,
the other two being the
Rhines and PV-mixing mechanisms already mentioned.
The Kelvin mechanism is simple to understand, especially
when the weak forcing is anisotropic in the same sense as
that of our injected vortex
pairs, with their east-west orientation.
Such pairs are immediately sheared into
phase-tilted structures producing upgradient Reynolds
stresses $\overline{\uu'\vv'}$.
The Taylor identity (\ref{eq:taylor-identity})
determines the accompanying $\overline{\vv'\qq'}$ field.
That field describes an eddy PV flux that
is upgradient
at some latitudes $y$
and downgradient at others
and,
as indicated by the
$y$ derivative in (\ref{eq:taylor-identity}),
involves subtle phase relations sensitive
to the $y$-gradients of disturbance amplitude
and shearing rates.

\begin{figure} [t]
\hspace{-0.50cm}
\includegraphics[width=22pc,angle=0]
{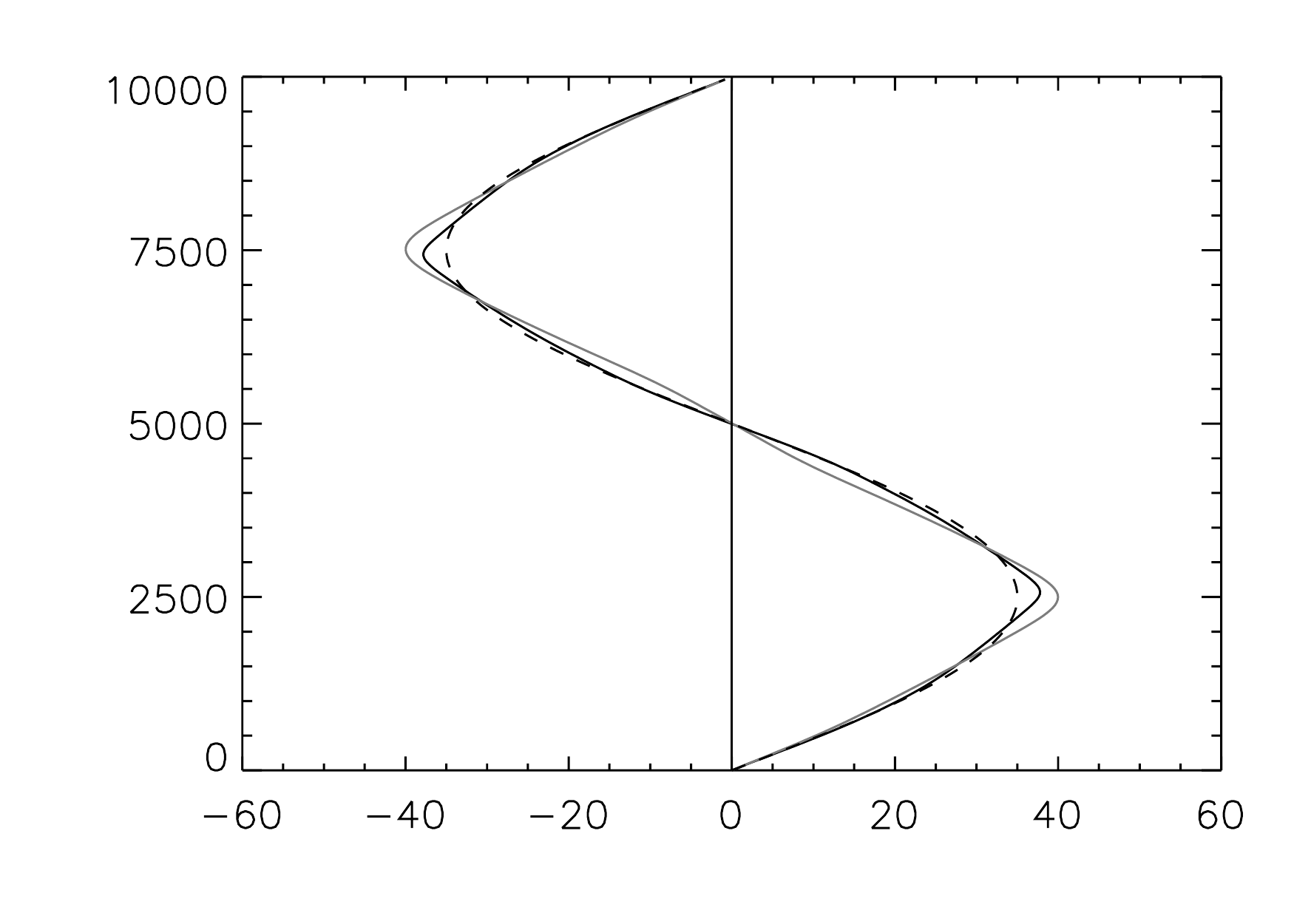}
\vspace{-1.2cm}
\caption{\footnotesize
Zonal-mean zonal velocity profiles $\uubar(y)$
in \ms\ ($y$ axis in \km)
for a pair of Kelvin-dominated, pure-DI runs
 \,\mbox{DI-12-0.5-64}\, and \,\mbox{DI-12-1-64}\, 
with $\LD=1200$\km\ and
$\qqstarmax=0.5$ (darker solid curve) and
$\qqstarmax=1.0$ (lighter solid curve),
with $\fractionalbiasmax=$ 1/64 and
with injection rates $\ttmax^{-1}$ increased by a factor 100,
as specified in the caption to Table~\ref{tab:parameter_values}.
Although it makes little difference to these profiles,
they have been time-averaged from $t=108$ to 202 Earth years
for consistency with the profiles of
$\zetabar$, $\qqbar$, and $\acforcing$ shown below,
some of which are more subject to fluctuations
within a statistically steady state.
The dashed curve is the
deep-jet velocity profile $\uudeep(y)-U_0$ as before.
}
\label{fig:kelvin-ubar-profiles}
\end{figure}

\begin{figure} [t]
\hspace{-0.50cm}
\includegraphics[width=22pc,angle=0]
{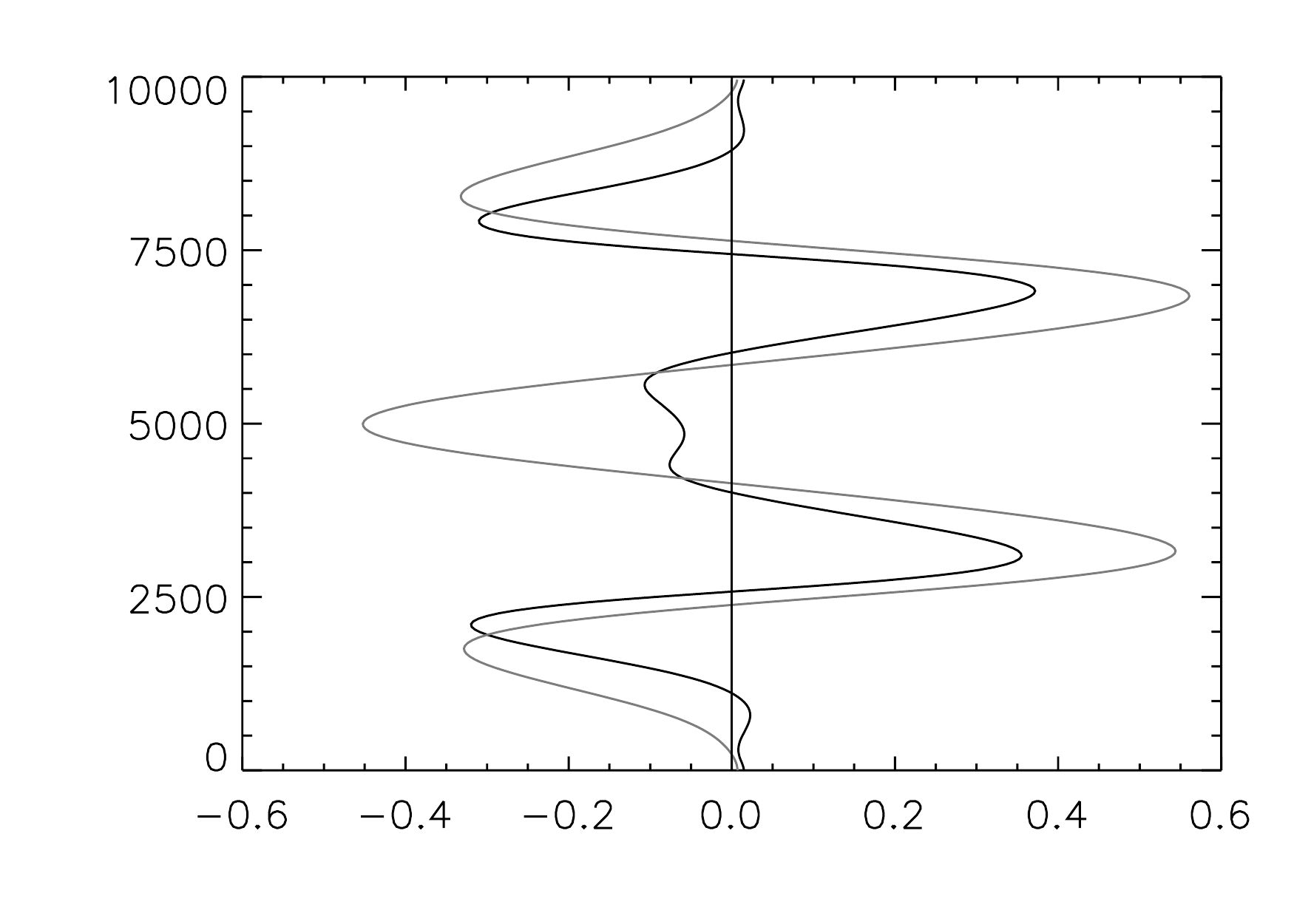}
\vspace{-1.2cm}
\caption{\footnotesize
Zonal-mean interface-elevation profiles
$\zetabar(y)$  ($y$ axis in \km)
for the same pair of Kelvin-dominated runs, with
$\qqstarmax=0.5$ (darker curve) and
$\qqstarmax=1.0$ (lighter curve),
both time-averaged from $t=108$ to 202 Earth years
as in Fig.~\ref{fig:kelvin-ubar-profiles}.
Without the time averaging, the $\qqstarmax=1.0$ curve would be
less symmetric and would fluctate noticeably,
because of
vacillations mentioned in the text.
}
\label{fig:kelvin-zetabar-profiles}
\end{figure}

\begin{figure} [t]
\hspace{-0.50cm}
\includegraphics[width=22pc,angle=0]
{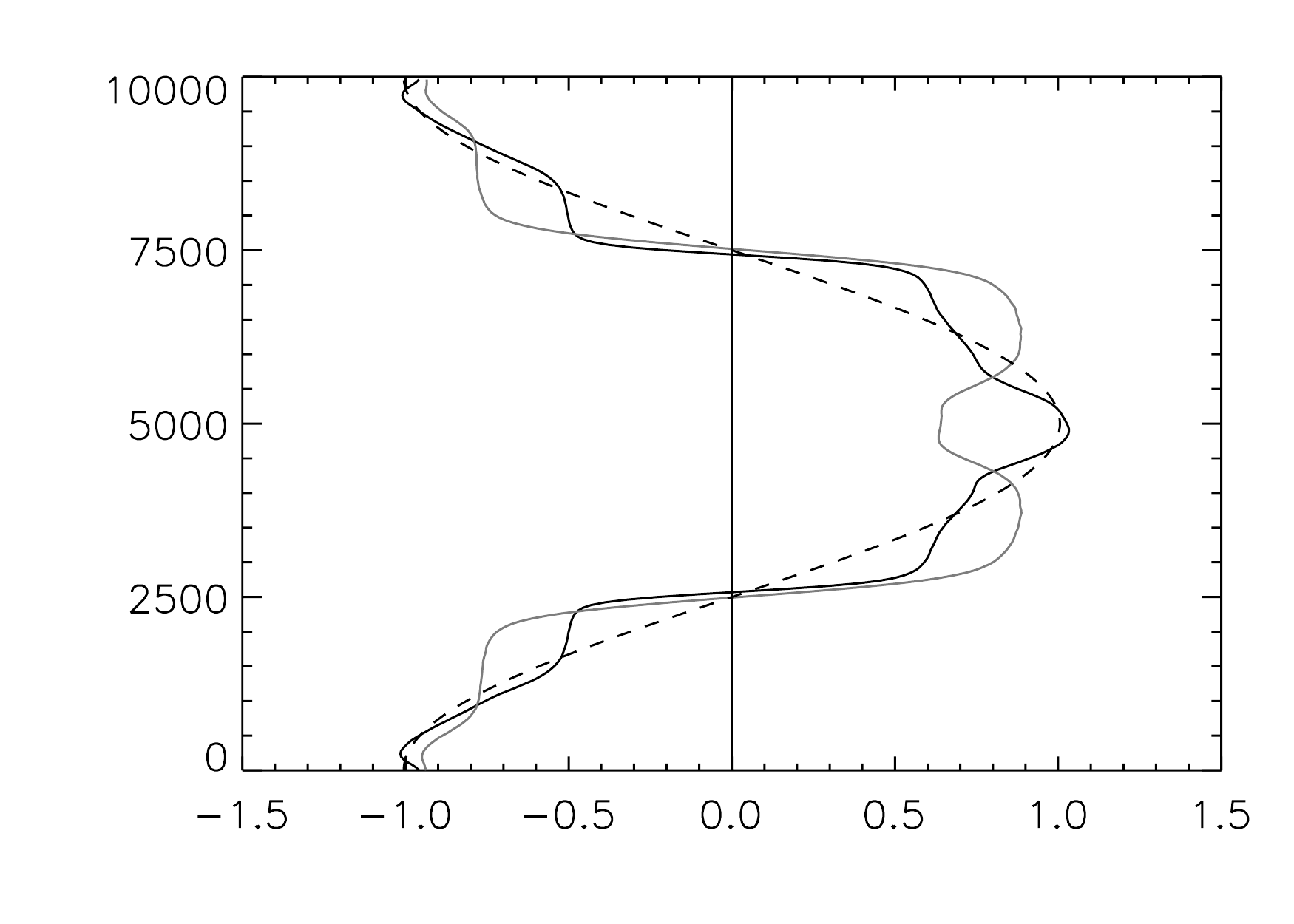}
\vspace{-1.2cm}
\caption{\footnotesize
Zonal-mean PV profiles $\qqbar(y)$ ($y$ axis in \km)
for the same pair of Kelvin-dominated runs, with
$\qqstarmax=0.5$ (darker solid curve) and
$\qqstarmax=1.0$ (lighter solid curve),
both time-averaged from $t=108$ to 202 Earth years
as in Fig.~\ref{fig:kelvin-ubar-profiles};
see text.
The dashed, sinusoidal curve is the initial PV profile.
The PV is in units of $\Umax/L=2.199\times10^{-5}$\smone\
as before.
}
\label{fig:kelvin-qbar-profiles}
\end{figure}

To suppress small-scale vortex activity and to allow the
Kelvin mechanism to dominate, we must take
$\qqstarmax$ small enough to ensure
that injections are almost always weak.
Figures~\ref{fig:kelvin-ubar-profiles}--\ref{fig:kelvin-qbar-profiles}
show statistically steady
$\uubar$, $\zetabar$, and $\qqbar$ profiles from a pair of
pure-DI runs with $\qqstarmax=0.5$ and~$1$
(darker and lighter curves respectively),
the runs labeled
 \,\mbox{DI-12-0.5-64}\, and \,\mbox{DI-12-1-64}.\, 
As before, we take $\LD=1200$\km.
The Kelvin mechanism is so weak that,
in order to see it working and to
reach statistical steadiness,
we had to reduce $\fractionalbiasmax$ to 1/64 and to increase the
average
injection rate by two orders of magnitude, i.e.,
we had to reduce $\ttmax$ by a factor 100,
as specified in the caption to Table~\ref{tab:parameter_values}.

\begin{figure} [t]
\hspace{-0.50cm}
\includegraphics[width=22pc,angle=0]
{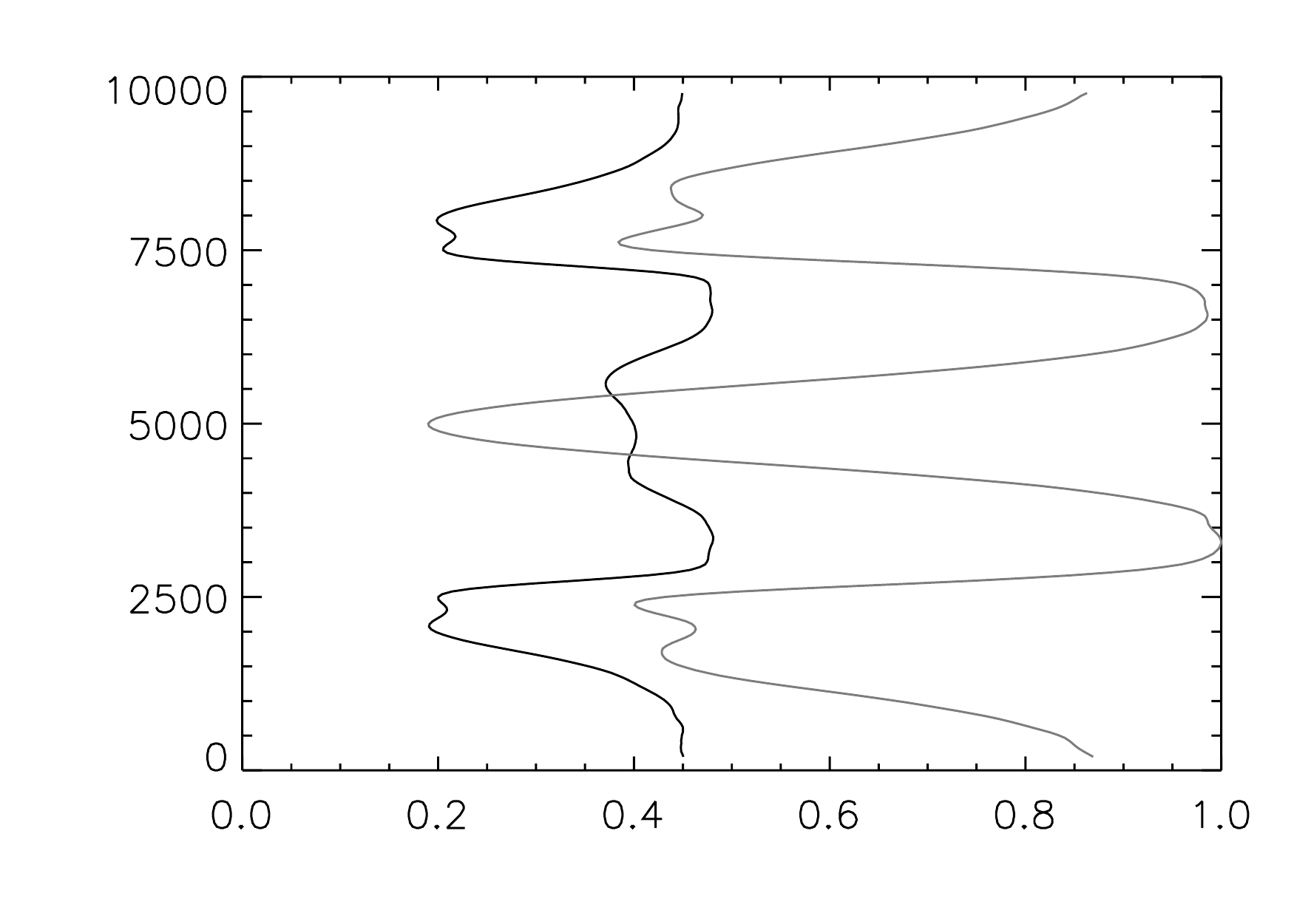}
\vspace{-1.2cm}
\caption{\footnotesize
Relative moist-convective activity profiles
$\acforcing(y)/\acforcingmax$ ($y$ axis in \km)
from (\ref{eq:acforcing-def})
for the same pair of Kelvin-dominated runs, with
$\qqstarmax=0.5$ (darker curve) and
$\qqstarmax=1.0$ (lighter curve),
both time-averaged from $t=108$ to 202 Earth years
as in Fig.~\ref{fig:kelvin-ubar-profiles}.
In the central part of the belt
convective activity is
inhibited, in both runs, for the reasons explained in the text.
}
\label{fig:kelvin-abar-profiles}
\end{figure}

The jets are indeed sharpened and the jet-core $\qqbar$ profiles
steepened, in both cases,
creating in turn a $\zetabar$ structure that is interesting but
unrealistic.  As Fig.~\ref{fig:kelvin-zetabar-profiles} shows,
the central part of the belt is
relatively warm, $\zetabar$ negative,
with only the edges cold,
$\zetabar$ positive.  The corresponding
$\acforcing$ profiles
from (\ref{eq:acforcing-def})
are given in
Fig.~\ref{fig:kelvin-abar-profiles},
showing an inhibition of convective activity in mid-belt.
No such inhibition is seen on the real planet.

The unrealistic $\zetabar$ structure arises from the way
the Kelvin mechanism works in a model with
no artificial Ray\-leigh friction, corresponding to the
low-friction limit found by \cite{Srinivasan2014},
their $\mu\to0$.  Each sheared vortex-pair
structure survives as long as it can, through nearly the whole
range of phase-tilt angles.
It is only when the orientation has become nearly
zonal
that the structure is destroyed by the model's
high-wavenumber filter.  Thus its lifetime
is inversely proportional to the local background shear
$\partial\uubar/\partial y$.
The Reynolds stress $\overline{\uu'\vv'}$, time-averaged over many
injections, is also, therefore, inversely proportional to
$\partial\uubar/\partial y$,
as in Eq.~(44) of \cite{Srinivasan2014}.
So as long as the $\uubar$ profile
remains close to its initially sinusoidal shape while the
$\zeta$ field
remains flat, keeping
injection strengths uniform everywhere,
the profile of $-\overline{\uu'\vv'}$ has a smooth,
positive-valued
U shape within
the belt, with a broad minimum in mid-belt and a steep increase
toward each jet extremum.  In the zone, with the sign of
$\partial\uubar/\partial y$
reversed,
it is
 $+\overline{\uu'\vv'}$
rather than
 $-\overline{\uu'\vv'}$
that is positive and
U-shaped; and there is a very steep transition
at each jet extremum producing a sharp, narrow peak in the zonal force
$-\partial(\overline{\uu'\vv'})/\partial y$,
positive at the prograde jet and negative at the retrograde.

The jet-sharpening is therefore strongly localized,
with a small $y$-scale $\ll\LD$.
It begins with narrow peaks
growing at the extrema of the otherwise-sinusoidal $\uubar$
profile, with
$|\partial\uubar/\partial y|$ reduced everywhere else.
The PV profile develops correspondingly sharp steps,
cut into the sides of its initially sinusoidal shape.
That is, there is localized jet-sharpening
but --- in striking contrast with
Fig.~\ref{fig:pureDI-ubar-profiles} ---
weakening rather than strengthening
of
$\uubar$ at
most other latitudes $y$.
The thermal-wind tilt of the interface is therefore,
at most other latitudes,
opposite to what it was in the cases discussed in
Section~\ref{sec:mainresults},
except within narrow regions near the jet peaks.
That is the essential reason why the belt develops a warm,
negative-$\zetabar$
central region with
cold, positive-$\zetabar$ regions only in the outer parts
of the belt.

The change in the $y$-profile of $\zetabar$
and hence of injection strengths
then reacts back on the $\overline{\uu'\vv'}$
profiles, but in a rather smooth way that leaves the
qualitative pattern unchanged.
Indeed, the back-reaction
acts as a positive feedback that \emph{reinforces}
the pattern, because the warm belt center weakens the injections
there and thus deepens the 
central minimum in the U-shaped profile of $-\overline{\uu'\vv'}$.
That is why the lighter curve in
Fig.~\ref{fig:kelvin-ubar-profiles}, corresponding to the
less weakly forced run, $\qqstarmax=1$,
shows a $\uubar$ profile
more conspicuously weakened across most of the belt.
The resulting thermal-wind tilt
further reinforces the
central warmth of the belt.

The stronger feedback for $\qqstarmax=1$
appears to be responsible for the central dip in
the $\qqbar$ profile
seen in the lighter curve
in Fig.~\ref{fig:kelvin-qbar-profiles}.
The central dip gives rise to
a weak long-wave shear instability
(though stronger when $\LD=1500$\km),
because
we then have nonmonotonic $\partial\qqbar/\partial y$
on a smaller $y$-scale, putting pairs of
counterpropagating Rossby waves within reach of each other.
This long-wave instability
produces vacillations in the form of
weak traveling undulations
with zonal wavenumber~1.
The vacillations hardly affect the
$\uubar$ and $\qqbar$ profiles,
but show up more clearly in a time sequence of
$\zetabar$ profiles.  The corresponding $\zetabar$ profile in
Fig~\ref{fig:kelvin-zetabar-profiles} (lighter curve)
has been time-averaged to reduce
the effects of these vacillations.

For $\qqstarmax=0.5$ we see a weak and
entirely different,
zonally symmetric, mode of
instability that causes spontaneous $y$-symmetry breaking
in the central region of the belt,
4000\km\ $\lesssim y \lesssim$ 6000\km.
For instance the $\uubar$ profile given by the darker
solid
curve in Fig.~\ref{fig:kelvin-ubar-profiles}
shows a tiny departure
from antisymmetry about mid-belt.
The darker
curves
in Figs.~\ref{fig:kelvin-zetabar-profiles}
and~\ref{fig:kelvin-abar-profiles}
are more conspicuously
asymmetric in the central region.
Considering a $\uubar$ profile consisting of a
constant cyclonic shear plus a small wavy
perturbation,
we see that such a perturbation is zonostrophically stable
--- because the abovementioned inverse proportionality
reduces $|\overline{\uu'\vv'}|$
wherever $|\partial\uubar/\partial y|$ increases ---
but ``thermostrophically unstable''
via the feedback from $\zetabar$,
which evidently has the opposite effect on
$|\overline{\uu'\vv'}|$ and predominates in this case.

\section{Concluding remarks}
\label{sec:conclu}

In view of the Kelvin regime's lack of realism we return to
the model's realistic, statistically steady, pure-DI
regimes that have been our main focus
(Sections~\ref{sec:mainresults}--\ref{sec:mechs-migration}).
In those regimes,
not only is $\overline{\uu'\vv'}$
persistently
upgradient,
but also $\overline{\vv'\qq'}$,
at nearly all latitudes $y$,
after sufficient time averaging.
The small-scale vortex activity
produces a persistent
migration of small anticyclones
from belts to zones.
It is this upgradient migration that \emph{strengthens}
the upper jets,
as distinct from the inhomogeneous PV mixing that \emph{sharpens}
them.  The importance of
migration on the real planet
was suggested by \citet{Ingersoll2000}.
In the model it is mediated by quasi-random walking
away from strong-injection sites,
via chaotic vortex interactions,
in combination with the
beta-drift mechanism
(Section~\ref{sec:mechs-migration} above).

Both mechanisms are entirely different from the Kelvin
jet-sharpening
mechanism because the latter
involves no vortex activity,
as already emphasized,
but only passive shearing of injected vortex pairs
by the background zonal flow $\uubar(y)$.  Passive shearing
of small-scale anomalies is also what seems to
produce the upgradient $\overline{\uu'\vv'}$
in the real planet's cloud-top winds
\citep[e.g.,][]{Salyk2006}.
Yet, in our model at least,
as shown in Section~\ref{sec:kelvin},
the Kelvin mechanism cannot
produce a realistic
$\zeta$ structure.
It therefore cannot, in this model, produce a realistic
belt-to-zone contrast in moist-convective activity.

We suggest therefore that the
cloud-top  $\overline{\uu'\vv'}$ on the real planet
must be a relatively shallow phenomenon, whose vertical scale is
much smaller
than the depth of the \wl.
It is most likely, we suggest, to result
not from the shearing of
tall, columnar vortices resembling
the injected vortices in our model
but, rather, from the
shearing of the real \whl's
small scale, baroclinic, fully three-dimensional fluid motions.
Such motions, including shallow vortices and the
real filamentary
moist convection
are, of course, outside the scope of any \oneandahalflayer\ model
and not simply related to PV fields like that of
Fig.~\ref{fig:pureDI-maps-zeta-and-pv}b above.

As is well known, the same conclusions are suggested by the
implausibly
large kinetic-energy conversion rates obtained when the
cloud-top  $\overline{\uu'\vv'}$ field is
assumed to extend downward,
along with $\partial\uubar/\partial y$,
throughout the entire \wl.
When one vertically integrates cloud-top conversion rates
$\overline{\uu'\vv'}\Sliver\partial\uubar/\partial y$,
whose global average $\sim10^{-4}$\wkg,
then global integration gives numbers
``in the range 4--8\percent\ of the total thermal energy emitted by Jupiter''
\citep[][\andrefs]{Salyk2006}.
Such large conversion rates are
overwhelmingly improbable, in a
low-Mach-number fluid system such as Jupiter's \wl.

Consistent with these considerations,
the model's flow re\-gimes with realistic $\zeta$ structures have
conversion rates
$\overline{\uu'\vv'}\Sliver\partial\uubar/\partial y$
that are still positive, but
about two orders of magnitude smaller.
For instance, in the case examined in
Section~\ref{sec:mainresults}\ref{subsec:realistic-example}
we find
$\overline{\uu'\vv'}\Sliver\partial\uubar/\partial y$
values that fluctuate around a time-mean close to
$1\times10^{-6}$\wkg\
\citep[][Fig.\ 4.13]{mythesis}.
Such values are much more plausible, for the whole \wl,
than the observed cloud-top values $\sim10^{-4}$\wkg.

As well as the model's success in producing flow re\-gimes
with realistically straight \whl\ jets and realistic,
internally generated
belt--zone contrasts in moist-convective activity,
we note again the implied restriction on $\LD$ values.  Such
a restriction holds in the model and also,
very probably, on the real planet.
In the model, we found that
realistic behavior requires $\LD\lesssim1500$\km\
at 35\degree\ latitude.
It therefore seems
probable
that values such as the
\mbox{5000\km} used
for all the midlatitude flow fields presented in
\citet{Li2006}, for instance,
are unrealistically large.

Perhaps the weakest aspect of the current model is the
artificial
condition (\ref{eq:saturation-constraint})
that we adopted in order
to avoid strong-cyclone runaway.
As is well known, the real planet's large cyclones
can be intensely convective,
presumably because they have
cold, high-$\zeta$ footprints.
No \oneandahalflayer\
vortex-injection scheme can come close to
representing the three-dimensional reality.
An attractive compromise, and a possible way of
dispensing with (\ref{eq:saturation-constraint}),
might be to introduce
an eddy viscosity whose value
intensifies
whenever and wherever the
model's convective activity intensifies.
This might capture some of
the dissipative effects of the
real, three-dimen\-sion\-al\-ly turbulent moist convection,
while still avoiding the use of Rayleigh friction
or other such artifice.

A localized, convection-dependent eddy viscosity
would have the advantage of, probably, allowing
realistic
statistically steady states with
a simpler
vortex-injection scheme, such as that described by
Eqs.~(\ref{eq:def-psi-lim})--(\ref{eq:anticyclones-before-saturation})
alone.  It could
automatically expand the core sizes,
and dilute the peak strengths, of
the strongest injected vortices
and thus prevent strong-cyclone runaway.
As an added bonus
it might even produce realistic
cases in which large anticyclones form
(cf.\ end of Section~\ref{sec:mechs-migration}).
Because the real planet's
large anticyclones are not ubiquitous,
there may be
a certain delicacy about the conditions that allow them to form.

Such questions must await future studies.
These could include
studies using general circulation models
in which the
lower boundary conditions,
about which there is so much uncertainty,
are replaced by conditions corresponding to
a flexible interface above prescribed deep jets.

\bigskip
\bigskip

\noindent\textbf{Acknowledgements:}
Emma Boland and Andy Thompson
kindly helped us with the model code.  We also thank them
for useful comments and and/or encouragement along with
James Cho,
Tim Dowling,
David Dritschel,
Leigh Fletcher,
Boris Galperin,
Thomas Gastine,
Gary Glatzmaier,
Peter Haynes,
Xianglei Huang,
Andy Ingersoll,
Chris Jones,
Kirill Kuzanyan,
Junjun Liu,
Inna Polichtchouk,
Peter Read,
Tapio Schneider,
Richard Scott,
Sushil Shetty,
Adam Showman,
Steve Tobias,
Richard Wood,
Bill Young,
and
Roland Young.
The paper's reviewers suggested many useful clarifications,
for which we are grateful.
We also thank the UK
Science and Technology Facilities Council
for financial support in the form of a research studentship.

\vbox to 1.2 cm{$\phantom{M}$}

\ifthenelse{\boolean{dc}}
{}
{\clearpage}
\bibliographystyle{ametsoc}
\bibliography{library}

\end{document}